\documentclass[preprint,12pt,authoryear]{elsarticle}

\usepackage{amssymb}
\usepackage{amsmath}
\usepackage{amsfonts}
\usepackage{bm}
\usepackage{array}
\usepackage{booktabs}
\usepackage{hyperref}
\usepackage{natbib}
\usepackage{algorithm}
\usepackage{algcompatible}
\usepackage{booktabs}
\usepackage{enumitem}
\usepackage{longtable}
\usepackage{array}
\usepackage{subcaption}
\usepackage{makecell}
\usepackage{xcolor}

\newcommand{\MBG}{MBG}
\newcommand{\Hyb}[1]{HYB$_{\text{#1}}$}
\newcommand{\psifield}{\psi}

\journal{Spatial Statistics}

\begin{document}

\begin{frontmatter}
\title{Decoupling Distance and Networks: Hybrid Graph Attention–Geostatistical Methods for Spatio-temporal Risk Mapping} 

\author[aff1]{Toba Temitope Bamidele\texorpdfstring{\corref{cor1}}{}}
\ead{btoba37@gmail.com}
\cortext[cor1]{Corresponding author}
\author[aff2]{Ezra Gayawan}
\author[aff2]{Femi Barnabas Adebola}
\author[aff3]{Olatunji Johnson}

\affiliation[aff1]{
  organization={Department of Statistics, Federal University of Technology and Environmental Sciences},
  city={Iyin Ekiti},
  state={Ekiti State},
  country={Nigeria}
}

\affiliation[aff2]{
  organization={Department of Statistics, Federal University of Technology},
  city={Akure},
  state={Ondo State},
  country={Nigeria}
}

\affiliation[aff3]{
  organization={Department of Mathematics, University of Manchester},
  city={Manchester},
  country={UK}
}

\begin{abstract}
Accurate spatial prediction and rigorous uncertainty quantification are central to modern spatial epidemiology and environmental risk analysis. We introduce a statistically principled hybrid modelling framework that constructs a covariate-similarity graph over the observation units, weights its edges by the attention coefficients of a dynamic Graph Attention Network (GATv2), and embeds the result as a structured latent effect fitted jointly with a latent Gaussian spatial process from model-based geostatistics (\MBG{}). This framework jointly captures relational dependence encoded in graph structures and continuous spatial dependence governed by physical proximity. We evaluate the proposed model via a simulation study and an applied analysis of malaria prevalence data, comparing its predictive accuracy, calibration, and uncertainty quantification against classical geostatistical models and residual kriging after the standalone GATv2 architectures. Our analyses show that residual kriging after  GATv2 does not adequately capture the structured spatial dependence, thereby limiting predictive accuracy. Conversely, geostatistical models provide effective spatial prediction through structured covariance functions yet are constrained by linear predictor assumptions and by their reliance on Euclidean distance to encode spatial structure. By integrating the attention-weighted graph with an explicit probabilistic spatial random field, the hybrid model captured the relational dependence, improved predictive accuracy wherever relational dependence is present, and provided more realistic uncertainty quantification in both simulation and applied settings. The gains were largest for calibration: in every simulation scenario with relational structure, coverage of nominal $95\%$ credible intervals for the latent risk rose from $0.50$--$0.66$ under \MBG{} to $0.93$--$0.94$, and weighted Brier scores improved by $21$--$27\%$. An ablation attributes this benefit to the graph structure: replacing the graph-structured effect with an unstructured one removes the gain entirely, and the non-Euclidean edges of the graph contribute to it while its geographic edges reduce it. Overall, the findings demonstrate that the hybrid model constitutes a statistically coherent and empirically robust framework for modelling complex spatial and spatio-temporal processes in settings where both distance-based and structure-based dependencies operate.
\end{abstract}

\end{frontmatter}

\section{Introduction}
Precision public health depends on the ability to critically estimate disease burden at fine spatial resolutions to guide targeted intervention strategies. This need is significant for vector-borne diseases such as malaria, where transmission is strongly influenced by heterogeneous environmental, ecological, and socio-demographic processes operating across space \citep{weiss2019mapping}. Model-Based Geostatistics (\MBG{}) has become the dominant statistical framework offering a principled approach to modelling spatial dependence through latent Gaussian random fields \citep{diggle2019model}. In practice, \MBG{} models are commonly implemented using Gaussian Processes approximated via the Integrated Nested Laplace Approximation combined with the Stochastic Partial Differential Equation (INLA-SPDE) approach \citep{lindgren2011explicit}. This approximation yields sparse precision matrices, enabling computationally efficient inference even for thousands of spatial locations while retaining the full probabilistic structure required for coherent prediction and uncertainty quantification. Nonetheless, standard \MBG{} formulations typically assume stationarity in the spatial process and linearity in the mean structure, which can limit their ability to represent complex, non-stationary disease dynamics. In many epidemiological systems, spatial correlation arises not only from geographic proximity but also from non-distance-based mechanisms such as ecological similarity, mobility networks, or shared environmental exposures. Classical \MBG{} models cannot represent such structure because of the reliance on Euclidean distance for the covariance functions. Machine learning methods, particularly graph neural networks, have the potential to represent these forms of dependence but lack principled methods for quantifying uncertainty.

Machine learning methods have gained prominence in disease mapping due to their capacity to model non-linear relationships and high-dimensional interactions without strong distributional assumptions. However, most conventional machine learning models lack explicit mechanisms to represent spatial dependence, often treating covariates derived from geographic coordinates as independent rather than as structured sources of correlation, leading to spatial overfitting and degraded predictive performance when extrapolating to unobserved locations \citep{meyer2019importance}. Recent methodological developments seek to reconcile these limitations by combining the probabilistic foundations of spatial statistics with the representational flexibility of modern machine learning, forming a hybrid framework. 

\cite{hengl2018random} introduced random forest for spatial prediction (RFsp), which incorporates spatial context by using distance-based covariates derived from geographic proximity to observation points. By encoding spatial relationships as informative features, RFsp relaxes the stationarity constraints inherent in traditional kriging and enables the modelling of complex, nonlinear interactions between environmental covariates and spatial location. \cite{chen2024deepkriging} proposed DeepKriging, which integrates spatial dependence by augmenting Deep Neural Networks (DNNs) with coordinate-based embedding layers constructed from spatial basis functions. This design circumvents the computational challenges of covariance-based geostatistical models while substantially enhancing the ability to model non-Gaussian, non-stationary, and highly nonlinear spatial processes by learning spatial representations jointly with covariate effects. \cite{chen2024dknn} addressed the interpretability limitations commonly associated with deep spatial models through the Deep Kriging Neural Network (DKNN). This framework combines a deep spatial encoder with a geostatistically inspired kriging decoder, enabling the model to produce spatial interpolation weights during prediction. In doing so, DKNN bridges the gap between the expressive power of deep learning and the transparency of classical geostatistics, offering improved interpretability without sacrificing predictive performance.
Although these methods explicitly incorporate spatial information, spatial dependence is still encoded indirectly either through engineered proximity-based covariates, fixed coordinate embeddings, or basis functions. As a result, spatial relationships are largely predefined rather than learned adaptively from data.

In this context, Graph Neural Networks (GNNs) have emerged as a promising framework for spatial modelling. By representing regions as nodes and spatial connectivity as edges, GNNs use message-passing protocols to aggregate information from local neighbourhoods, thereby explicitly encoding relational structure \citep{wu2020comprehensive}. Graph Attention Networks (GATs) further refine this approach by employing self-attention mechanisms to learn adaptive, data-driven weights for neighbouring nodes. Unlike standard convolution, this allows the model to capture directional relationships, acknowledging that the influence of neighbours varies across the domain \citep{velivckovic2017graph}. Building on this, recent advancements have introduced dynamic variants of GATs (GATv2) that address the GAT architecture's "static attention" limitation. By using a strictly dynamic attention mechanism, GATv2 ensures that attention weights are conditioned on both the source and target nodes, thereby offering greater expressive power for modelling complex spatial interactions \citep{brody2021attentive}. 

Functionally, this attention mechanism operates as a dynamic, data-driven spatial weighting scheme. By computing attention coefficients from interactions between nodal features, the model inherently accommodates non-stationarity and local directional effects \citep{brody2021attentive}. This mechanism learns a flexible spatial structure in which the strength of connectivity varies across the domain, rather than imposing a global, invariant rule for spatial decay. This allows the model to define the effective spatial neighbourhood not merely by geometric distance or topological adjacency, but also by the conditional similarity of the underlying processes. Two distinct roles are involved here, and it is useful to separate them. Graph construction determines \emph{which} units are treated as neighbours, while the attention mechanism learns \emph{how strongly} each neighbour contributes, conditional on that fixed edge set. The results reported below indicate that the first affects predictive performance; they do not establish how much the second adds once the first is in place.

Motivated by recent advances in GNNs and the persistent limitations of distance-based spatial dependence in existing geostatistical and hybrid models, this study proposes a hybrid framework that represents relational dependence as a graph-structured latent Gaussian effect, built on a covariate-similarity neighbourhood and weighted by GATv2 attention, and fits it jointly with explicit distance-based spatial dependence. By coupling the attention-weighted graph with the geostatistical latent process, the proposed model enables adaptive and flexible modelling of spatial dependence while preserving the inferential rigour and principled uncertainty quantification characteristic of model-based geostatistics.

This study makes three contributions. First, it shows that a graph-structured latent effect built from learned relational similarity improves both prediction and calibration where non-Euclidean dependence is present, and has little effect where it is not. Second, it examines which part of the attention mechanism led to the improvement, showing that the non-Euclidean component of the graph contributes while its geographic component does not. It also indicates which comparisons the experiments can and cannot separate. Third, it assesses whether the graph-based and distance-based effects are separately identified in these settings, and describes the conditions under which they are not. The methodology is evaluated through a simulation study and applied to real-world malaria prevalence data from Nigeria, demonstrating improved predictive accuracy and spatial generalisation relative to conventional geostatistical benchmarks.

\section{Methods}
\subsection{Model-Based Geostatistics (\MBG{})}
Let $Z_i$ denote the observed number of cases among $n_i$ trials at spatial location $s_i \in \mathbb{R}^2$ and time $t_i \in \mathbb{R}$ or $\{1,\ldots,T\}$, for $i=1,\ldots,m$. We assume a binomial observation model,

\begin{equation}
Z_i \mid p_i, n_i \sim \operatorname{Binomial}(n_i,p_i),
\label{eq:mbg_binomial}
\end{equation}
where $p_i$ denotes the underlying prevalence or probability of a case at location $s_i$ and time $t_i$. The probability is linked to the latent spatio-temporal process through the logit link,

\begin{equation}
\operatorname{logit}(p_i)
= \beta_0 + \mathbf{x}_i^\top\boldsymbol{\beta} +
S(s_i,t_i)+ \varepsilon_i, \qquad \varepsilon_i \stackrel{\text{iid}}{\sim} N(0,\tau^2).
\label{eq:mbg_model}
\end{equation}
Here, $\beta_0$ is the intercept, $\mathbf{x}_i^\top\boldsymbol{\beta}$ is the fixed-effect component, $\mathbf{x}_i$ is a vector of covariates with corresponding regression coefficients $\boldsymbol\beta$ and $S(s_i,t_i)$ is a zero-mean spatio-temporal Gaussian random field that captures residual spatial and temporal dependence, and $\varepsilon_i$ is the noise term called the nugget effect.

For computationally efficient Bayesian inference, the spatio-temporal Gaussian random field $S(s_i,t_i)$ is estimated using the stochastic partial differential equation (SPDE) approach, which approximates the continuously indexed Mat\'ern Gaussian field by a Gaussian Markov random field defined on a triangulated spatial mesh, resulting in sparse precision matrices \citep{lindgren2011explicit}. Inference is performed using the Integrated Nested Laplace Approximation (INLA), a deterministic algorithm designed for fast, approximate Bayesian inference \citep{rue2009approximate}.

\subsection{Spatio-Temporal Residual Kriging}
Spatio-temporal residual kriging is used to account for spatial and temporal dependence that remains after the GATv2 model has explained the variation associated with the covariates. The GATv2 model described in Section~\ref{sec:GATv2_section} provides the model-based component of the prevalence surface. Spatio-temporal residual kriging is then used to account for spatial and temporal dependence that remains after the GATv2 predictions. Assuming a binomial response, $Z_i \mid p_i,n_i \sim \mathrm{Binomial}(n_i,p_i)$, and $\operatorname{logit}(p_i) = \eta_i$, the underlying process is defined as

\begin{equation}
\eta(s_i,t_i) = \hat{\eta}_{GATv2}(s_i,t_i) + \upsilon(s_i,t_i),
\end{equation}
where $\hat{\eta}_{GATv2}(s_i,t_i)$ denotes the prediction from the GATv2 model and $\upsilon(s_i,t_i)$ represents the residual component with remaining spatio-temporal dependence. 

The representation of the residuals depends on the observation structure. When the same $n$ locations are observed at each of $m$ time points, the observations form a complete space-time grid of $n\times m$ location-time combinations, which are stored as a Space Time Full Data Frame (STFDF). However, when observations are available at different locations and times, without a complete location-time grid, they are instead stored as a Space Time Irregular Data Frame (STIDF) object. In this case, each record contains an observed spatial location, time, and residual, without requiring every possible location-time combination to be observed \citep{pebesma2015package}. 

The observed residuals are used to estimate the empirical spatio-temporal dependence structure, and a parametric variogram model is fitted to this structure. The fitted variogram is then used to predict the residual component at unobserved space-time locations. The final prediction therefore adds the GATv2 prediction at unobserved locations with the kriged residual. This allows the GATv2 model to provide the primary prediction while residual kriging accounts for remaining spatio-temporal dependence that is not captured by the learned model.

\subsection{Dynamic Graph Attention Network (GATv2)}\label{sec:GATv2_section}
GATv2 is a neural network model designed to analyse data naturally represented as a graph of nodes and edges. Graph-structured data arises naturally in many real-world problems. In such data, observations are not independent; instead, each node interacts with its neighbours through edges. GATv2 handles situations where the connections between data points are just as meaningful as the data itself. For example, in modelling disease transmission, a village's infection risk may depend on its connections to other villages. For vector-borne diseases such as malaria, this spatial interdependence is driven by vector ecology and movement. Villages sharing river systems or located within mosquito flight ranges, which vary by species from 50 m to 50 km, may experience linked infection dynamics \citep{VERDONSCHOT201469}. Consequently, a village’s risk profile is shaped by factors beyond node features or distance to other nodes. To capture these dependencies, GATv2 learns how to combine these features by focusing on the most relevant neighbours. It does this through a mechanism called 'attention', which assigns different weights to neighbours, allowing the model to focus more on the most influential ones. These neighbours are determined by a predefined graph structure, where the edges specify which nodes are connected. GAT learns which neighbours matter most for each node, becoming highly effective in complex domains where relationships vary in strength and importance.

Let $\mathcal{G}=(\mathcal{V},\mathcal{E})$ be the input graph with $N=|\mathcal{V}|$ nodes and initial node features ${H}^{(0)}\in\mathbb{R}^{N\times d_0}$. Each node represents a spatial observation at a given time step, while the edge set $\mathcal{E}$ is constructed prior to model training using predefined neighbourhood criteria. A GATv2 layer $l$ with $K$ attention heads transforms ${H}^{(l)}$ into ${H}^{(l+1)}$ as follows.

For each head $k=1,\dots,K$ and every ordered pair $(i,j)$ with $\mathcal{N}(i)=\{j:(i,j)\in\mathcal{E}\}\cup\{i\}$, we compute:

\begin{equation}
e_{ij}^{(l,k)} = \left( {a}^{(l,k)} \right)^\top \mathrm{LeakyReLU}\left( {W}^{(l,k)} \begin{bmatrix} {h}_i^{(l)} \\ {h}_j^{(l)} \end{bmatrix} \right) \in \mathbb{R}
\label{eq:gatv2_z}
\end{equation}

\noindent where ${W}^{(l,k)}$ and ${a}^{(l,k)}$ are learnable parameters, ${h}^{(l)}_i \in \mathbb{R}^{d_l}$ is a feature vector of node $i$ at layer $l$, corresponding to the $i$-th row of $H^{(l)}$, $d_l$ is the dimensionality of node features at layer $l$, $e^{(l,k)}_{ij} \in \mathbb{R}$ is the unnormalized attention coefficient measuring the importance of node $j$'s features to node $i$ under attention head $k$ at layer $l$.
The crucial modification over the original GAT is the application of LeakyReLU before the dot product with ${a}^{(l,k)}$, which renders the attention mechanism dynamic \citep[Theorem~1]{brody2021attentive}.

The attention coefficients determine how much information node $i$ receives from each of its neighbouring nodes. For each attention head $k$, the unnormalised attention score $e_{ij}^{(l,k)}$ is converted into a weight using a softmax over the neighbourhood of node $i$:

\begin{equation}
\alpha_{ij}^{(l,k)} = \frac{\exp(e_{ij}^{(l,k)})}{\sum_{m\in\mathcal{N}(i)} \exp(e_{im}^{(l,k)})}
\label{eq:gatv2_alpha}
\end{equation}

The resulting coefficients are non-negative and sum to one across the neighbours of node $i$. Thus, a larger $\alpha_{ij}^{(l,k)}$ indicates that node $i$ receives greater weight when information from the neighbourhood is used to update node $i$. The softmax therefore converts the learned pairwise scores into relative weights that determine how information is shared between connected observations.

For each attention head, the representation of each neighbouring node is first transformed using a separate value projection matrix ${V}^{(l,k)}\in\mathbb{R}^{d_{\text{out}}^{(l)} \times d^{(l)}}$. The transformed representations are then weighted by the attention coefficients and summed to obtain the updated representation of node $i$:

\begin{equation}
{h}_i^{(l+1)} = \underset{k=1}{\overset{K}{\parallel}} \; \sigma \bigg(\sum_{j\in\mathcal{N}(i)}\alpha_{ij}^{(l,k)} \, {V}^{(l,k)} {h}_j^{(l)}\bigg).
\label{eq:gatv2_head}
\end{equation}

For the intermediate layers, the $K$ attention heads operate independently, and their outputs are concatenated. The resulting representation is then passed through the non-linear activation function $\sigma$. Each attention head can therefore learn a different weighting of the neighbouring observations, allowing the model to represent different patterns of relational dependence.

In the final GATv2 layer, the outputs of the $K$ attention heads were averaged rather than concatenated. No non-linear activation is applied separately to each head before averaging. The averaged representation is then passed through the activation function:

\begin{equation}
{h}_i^{\text{final}} = \sigma\bigg(\frac{1}{K} \sum_{k=1}^{K} \sum_{j\in\mathcal{N}(i)}\alpha_{ij}^{(l,k)} \, {V}^{(l,k)} {h}_j^{(l)}\bigg).
\label{eq:gatv2_final}
\end{equation}

Averaging the final attention heads produces a single representation for each node while retaining information learned from the different attention heads. For node-level regression, the model's final predicted value for node $i$, given as $\hat{O}_i$, is:

\begin{equation}
\hat{O}_i = {w}_{\text{out}}^\top {h}_i^{\text{final}} + b_{\text{out}} \in \mathbb{R}
\label{eq:node_pred}
\end{equation}
where $b_{\text{out}}$ is a learnable scalar bias term, $w_{\mathrm{out}}$ is the learnable output weight vector, and $h_i^{\mathrm{final}}$ is the final node representation produced by the GATv2 network. 

\subsection{GATv2 Model Specification}
\subsubsection{Network Target and Pre-whitening} 
Given the binomial model in equation \ref{eq:mbg_binomial}, the GATv2 network is not required to model $\operatorname{logit}(p_i)$ in full. Instead, the linear covariate effect is first removed through a pre-whitening step, so that GATv2 can model only the part of the signal the linear terms cannot represent. Concretely, we form the empirical logit  with a continuity correction,
\begin{equation}
  \tilde\eta_i
  = \log\!\frac{Z_i + \tfrac12}{\,n_i - Z_i + \tfrac12\,},
\end{equation}
where the $\tfrac12$ ensures that $\tilde\eta_i$ remains finite when $Z_i = 0$ or $Z_i = n_i$. This is necessary because prevalence spans $0.01$ to $1.00$ across our scenarios. We regress $\tilde\eta_i$ on the covariates using the training units only, obtaining a linear predictor $\hat\ell_i = \mathbf{x}_i^{\top}\hat{\boldsymbol\beta}^{\text{pre}}$.
The pre-whitening serves only to give the GATv2 network a target free of the linear trend.

The pre-whitening predictor $\hat{\ell}_i$ could alternatively be obtained by
fitting a binomial Generalised Linear Model (GLM) rather than by least squares to the empirical logit $\tilde{\eta}_i$. We compared these two approaches in the simulation study and
found that their resulting linear predictors were highly correlated ($r=0.999$). However, the binomial GLM coefficients were attenuated by approximately 10--20\% relative to the least-squares estimates and were further from the true data-generating coefficients. This difference is expected because the pre-whitening model deliberately excludes the latent effects and nonlinear covariate effects. Under such unmodelled heterogeneity, logistic-regression coefficients can be attenuated towards zero, whereas least-squares fitting on the logit scale preserves the additive structure of the data-generating process. We therefore use least-squares pre-whitening as the primary specification.

\subsubsection{Loss Function and Network Training} 
Let $f(\mathbf{x}_i)$ be the GATv2 output. $f(\mathbf{x}_i)$ enters $p_i=1\big/(1+\exp(-\eta_i))$ in the loss function as $\eta_i = \hat\ell_i + f(\mathbf{x}_i)$. We considered two alternative loss functions for training the GATv2 network: a binomial negative log-likelihood (Binomial NLL) and mean squared error (MSE). Both objectives were evaluated within the simulation study to assess the extent to which the choice of training criterion affects predictive performance. The comparative analysis (Table~\ref{tab:loss_fxn_compared}) demonstrates that Binomial NLL yields slightly better predictive performance than MSE, justifying its selection for the model.

For $Z_i$ observed cases among $n_i$ trials, the network is trained by minimising the Binomial NLL:
\begin{equation}
\mathcal{L}_{\mathrm{Bin}}
= -\frac{1}{\sum_{i \in \text{train}} n_i}\sum_{i \in \text{train}}
   \bigl\{ Z_i \log p_i + (n_i - Z_i)\log (1- p_i)\bigr\}.
\label{eq:binomial_loss}
\end{equation}
This matches the likelihood for $Z_i$ used to generate the data and therefore accounts for the differing precision of the observed prevalence across units. In particular, the approximate variance of the empirical logit satisfies 
\[\operatorname{Var}\{\operatorname{logit}(\frac{Z_i}{n_i})\} \approx \frac{1}{n_i p_i(1-p_i)}\] so the amount of information contributed by each unit can vary by orders of magnitude as $n_i$ and $p_i$ vary. This heterogeneity is not adequately captured by treating the observed prevalence as equally precise responses under MSE. 

For generalisation to other response distributions, the released code provides implementations of MSE loss, denominator-weighted loss, and Binomial negative log-likelihood (NLL), enabling the framework to accommodate a broader range of response distributions.

\subsubsection{Graph Construction and Edge Definition}
The graph edges are formed from three optional sources, combined by union. Let $t(i)$ denote the time index of node $i$.

\begin{enumerate}
\item Covariate-similarity edges: Within each time slice,  each node is joined to its $k_{features} = 5$ nearest neighbours in standardised covariate space under Euclidean distance. To examine how the choice of covariates used to construct the similarity network affects the results, two specifications of the covariate-similarity edges were considered for the model. In the first specification, all the covariates were used to compute the pairwise distances and construct the nearest-neighbour network. In the second specification, the network was constructed using a single influential covariate. This second specification assesses whether a similarity network based specifically on the most influential covariate corresponding to hypothesised drivers of connectivity, such as human mobility, hydrological connectivity, or other relevant domain-specific processes, produces different results from one based on the full covariate set.
\item Spatial edges: Within each time slice, each node is joined to its $k_{spatial} = 5$ nearest neighbours in geographic space.
\item Temporal edges: Temporal edges connect nodes across successive time steps up to a predefined maximum lag, with the number of neighbours decreasing as the temporal lag increases. For lags $\ell = 1, \dots, L$ with $L = 2$, each node at time $t$ is joined to its $\max(1, k_{time} - \ell + 1)$ geographically nearest nodes at time $t + \ell$, with $k_{time} = 3$.
\end{enumerate}

Covariate and spatial edges are symmetrised ($i \sim j$ if either is among the other's $k$ nearest neighbours), so the graph is undirected. Temporal edges are formed forward in time. Self-loops are excluded at construction. To evaluate the contribution of each neighbourhood definition, graphs are constructed using different combinations of these edge types (Table \ref{tab:models}). The resulting edge set defines the neighbourhood over which GATv2 performs message passing and computes attention. The graph is built over all nodes required for both fitting and prediction, so a prediction location enters the graph exactly as an observed one does: its covariates and coordinates determine its neighbours, and it participates in message passing. What distinguishes it is that its \emph{response} is withheld (set to missing) and it contributes nothing to the training loss.

\subsection{Attention-Based Stochastic Process}
To examine the structural relations captured by GATv2, a weighted graph Laplacian was derived from the model’s learned attention coefficients. These coefficients provide data-adaptive weights describing the relative contribution of neighbouring nodes during message passing \citep{kim2022neuralprocessesstochasticattention}.  This procedure transforms the directed and multi-head attention outputs into a symmetric representation that encodes the strength of pairwise interactions between nodes.

The attention coefficients $(\alpha_{ij}^{(h)})$ were obtained from the final  GATv2 layer for each attention head $h \in \{1, \dots, H\}$ and directed edge $(i \rightarrow j)$.  To produce a single attention value per edge, we computed the mean attention across all heads $\bar{\alpha}_{ij} = \frac{1}{H}\sum_{h=1}^{H} \alpha_{ij}^{(h)}$. These averaged coefficients were used to define a directed weighted adjacency mapping. $W_{ij} = \bar{\alpha}_{ij}$.  To analyse undirected relationships, the weights were symmetrised by averaging the coefficients in both directions $W^{\text{sym}}_{ij} = W^{\text{sym}}_{ji} = \frac{1}{2}\bigl(W_{ij} + W_{ji}\bigr)$. The symmetric adjacency weight thus represents the mutual intensity of attention shared between node pairs.

The corresponding node degree for each vertex $i$ is defined as the sum of its symmetric incident weights $d_i = \sum_{j} W^{\text{sym}}_{ij}$. From these degrees and edge weights, we formed the unnormalized graph Laplacian $L = D - W^{\text{sym}}$, where $D = \mathrm{diag}(d_1, d_2, \dots, d_N)$ denotes the diagonal degree matrix \citep{kipf2016semi}.  The Laplacian entries are therefore given by
\begin{equation}
L_{ij} =
\begin{cases}
d_i, & i = j, \\
-\,W^{\text{sym}}_{ij}, & i \neq j.
\end{cases}
\label{eq:laplacian}
\end{equation}

To obtain numerically stable magnitudes for subsequent analysis, the Laplacian 
was scaled by the median node degree:
\begin{equation}
    C = \frac{1}{\mathrm{median}(d_i)}L
\label{eq: scaled laplacian}
\end{equation}

The attention-derived structure matrix $C$ induces a dependency pattern learned from graph attention weights rather than physical proximity or Euclidean distance. The scaling preserves the relative connectivity structure encoded by the attention-induced graph while standardising its overall magnitude.

This structure matrix is then supplied to the \textit{Type-I Generic model} implemented in the INLA framework, which for a given structure matrix $M$ is defined as $Q = \tau\{I - (\beta/\lambda_{\max}(M))\,M\}$, where $\lambda_{\max}(M)$ is the largest eigenvalue of $M$. Because $C$ is a graph Laplacian, its off-diagonal entries are already negative, so supplying $C$ itself would give a precision matrix with positive off-diagonal entries and therefore negative conditional correlation between graph neighbours, which is the opposite of the intended sharing of information. We therefore supply the complement
\begin{equation}
    C^{\star} = \lambda_{\max} I - C ,
\label{eq:cstar}
\end{equation}
where $\lambda_{\max}$ is the largest eigenvalue of $C$. Because $C$ is a Laplacian its smallest eigenvalue is zero, so $\lambda_{\max}(C^{\star}) = \lambda_{\max}(C) = \lambda_{\max}$, and $C^{\star}$ is positive semi-definite with the same sparsity pattern as $C$. The resulting precision matrix is
\begin{equation}
Q(\tau,\beta)=
\tau\left(
I-\frac{\beta}{\lambda_{\max}}C^{\star}
\right)
=
\tau\left[
(1-\beta)\,I+\frac{\beta}{\lambda_{\max}}\,C
\right],
\label{eq:precision_matrix}
\end{equation}
where $I$ is an identity matrix. Equation~\eqref{eq:precision_matrix} is a convex combination of an independent node effect and a Laplacian-smoothed effect. It has the form of the mixed model for spatial dependence proposed by \citet{leroux2000estimation}, applied here to a learned attention Laplacian in place of a fixed spatial adjacency, and it is closely related to the scaled mixing parameterisation used in disease mapping by \citet{riebler2016intuitive}. The scalar parameter $\tau$ governs the precision of the latent process, while $\beta \in [0,1)$ controls the strength of dependence encoded by $C$, with $\beta = 0$ giving an independent node effect and $\beta \rightarrow 1$ approaching smoothing driven entirely by the attention graph. Because $Q$ depends on $C$ only through $C/\lambda_{\max}$, the model is invariant to the rescaling in Eq.~\eqref{eq: scaled laplacian}; that rescaling is retained because it keeps the exported matrices on a comparable numerical scale across fits.

The precision matrix $Q(\tau,\beta)$ defines the conditional dependence structure of the latent field. In particular, nonzero off-diagonal entries correspond to conditional dependencies between connected nodes, while zero entries imply conditional independence given the remaining nodes. This matrix defines the structure of the random effect component referred to as the \textit{attention-based stochastic process} $A(s_i,t_i)$,

\begin{equation}
\label{eq:attention_effect}
A(s_i,t_i)\mid\tau,\beta
\sim
\mathcal{N}\left(
0,
Q(\tau,\beta)^{-1}
\right),
\}
\end{equation}
where $A(s_i,t_i)$ denotes the latent attention-based effect at location $s_i$ and time $t_i$. Embedding the attention-derived structure matrix within the Type-I INLA generic prior produces a Gaussian Markov random field whose local conditional dependencies mirror the attention-weighted relationships among spatial units. Consequently, the model captures latent similarity structures driven by ecological, demographic, mobility-related, or other domain-specific processes represented in the graph, rather than relying solely on physical adjacency.

To facilitate Bayesian inference and enable interpretable prior specification, 
the hyperparameters $(\tau, \beta)$ are reparameterized on unconstrained scales as $\theta_1 = \log(\tau)$, $\theta_2 = \mathrm{logit}(\beta)$. This transformation guarantees that $\tau > 0$ and $\beta \in (0,1)$, 
while allowing prior distributions to be defined directly on $(\theta_1, \theta_2)$, thereby improving numerical stability and providing flexibility when controlling the influence of priors on the variance and dependency strength.

\begin{algorithm}[htbp]
\caption{From fitted GATv2 attention to the precision matrix
$Q(\tau,\beta)$}
\label{alg:C}
\begin{algorithmic}[1]
\REQUIRE trained GATv2; node features $X$; edge set $E$; MC-dropout
  samples $S$
\State \textbf{MC-averaged attention.} For $s = 1,\dots, S$ run a forward
  pass with dropout active, extract the attention coefficients of the
  second GATv2 layer, and average over heads. Average over the $S$
  passes to obtain one weight $\alpha_{ij} \ge 0$ per directed edge
  $(i,j) \in E$.
\State \textbf{Symmetrise.}
  $w_{ij} = \tfrac{1}{2}(\alpha_{ij} + \alpha_{ji})$ for $i \neq j$,
  taking $\alpha_{ij} = 0$ where $(i,j) \notin E$. Set $w_{ii} = 0$.
\State \textbf{Laplacian.} $D = \operatorname{diag}(d_1,\dots,d_N)$ with
  $d_i = \sum_{j} w_{ij}$; set $L = D - W$.
\State \textbf{Rescale.} $C = L / \operatorname{median}\{d_i : d_i > 0\}$,
  so that the exported matrices are on a comparable numerical scale
  across fits.
\State \textbf{Complement.} $C^{\star} = \lambda_{\max} I - C$, with
  $\lambda_{\max}$ the largest eigenvalue of $C$. This is the matrix
  passed to \texttt{generic1}, and $\lambda_{\max}(C^{\star}) = \lambda_{\max}$.
\State \textbf{Precision.}
  $Q(\tau,\beta) = \tau\!\left(I - \dfrac{\beta}{\lambda_{\max}}C^{\star}\right)
   = \tau\!\left[(1-\beta)I + \dfrac{\beta}{\lambda_{\max}}C\right]$,
  with $\beta \in [0,1)$.
\end{algorithmic}
\end{algorithm}

\subsubsection{Properties of the Effect}
By construction, $C$ is symmetric with non-negative eigenvalues and zero row sums (verified numerically: $\max_i |\sum_j C_{ij}| < 10^{-8}$). Since $0 \le \lambda(C) \le
\lambda_{\max}$, the eigenvalues of $(1-\beta)I + (\beta/\lambda_{\max})C$ lie in
$[1-\beta,\,1]$, so $Q$ is positive definite for every $\beta \in [0,1)$ and $\tau > 0$; no further constraint is needed. The associated quadratic form is
\[
  A^{\top} Q A
  = \tau\left[
      (1-\beta)\sum_i A_i^2
      + \frac{\beta}{\lambda_{\max}}
        \sum_{i<j} \tilde{w}_{ij}\,(A_i - A_j)^2
    \right],
\]
where $\tilde{w}_{ij}$ denotes the off-diagonal magnitudes of $C$, that is, the attention weights after rescaling. The second term penalises differences between nodes joined by a strong attention edge, so large $\beta$ shrinks such nodes towards one another. Note that $Q \succeq \tau(1-\beta)I$, so the marginal variance of $A_i$ is at most $\{\tau(1-\beta)\}^{-1}$ and $\tau^{-1/2}$ should not be read directly as the marginal standard deviation of the field.

Writing $A \sim N(0, Q^{-1})$, the conditional expectation of $A_i$ given all other components of $A$ is
\[
  \mathbb{E}(A_i \mid A_{-i})
  = \frac{\beta \sum_{j \neq i} \tilde{w}_{ij} A_j}
         {(1-\beta)\lambda_{\max} + \beta \tilde{d}_i} ,
\]
where $\tilde{d}_i$ is the corresponding diagonal entry of $C$. This is a weighted average of neighbours in which the weights are the learned attention coefficients. $\beta$ controls the strength of dependence: $\beta \to 0$ gives independent effects, $\beta \to 1$ maximal smoothing over the attention graph. This is the sense in which $A$ is an \emph{attention-weighted} spatial-analogue effect: units that the network judged mutually informative are shrunk towards one another, irrespective of geographic distance.

\textbf{Relation to the CAR family.} If the $w_{ij}$ were binary and the rescaling omitted, Eq.~\eqref{eq:precision_matrix} would reduce exactly to the mixed spatial model of \citet{leroux2000estimation}, in which $\beta$ mixes an independent effect with an intrinsic Gaussian Markov random field defined by a fixed adjacency \citep{besag1991bayesian,rue2005gaussian}. The proposed construction differs only in that the weights come from a fitted attention mechanism rather than from a fixed adjacency, so the neighbourhood structure is learned from the data rather than specified in advance.

\subsection{The Hybrid Framework}
To address the limitations of traditional linearity and stationarity assumptions, as well as the restriction to Euclidean distances in modeling spatial dependence, while preserving robust uncertainty quantification, we propose a hybrid model that integrates the linear effects of the covariates, the nonlinear predictive capability of GATv2 \textit{$\hat{O_i}$} (Eq.~\eqref{eq:node_pred}), an attention-based stochastic process \textit{$A(s_i, t_i$)} (Eq.~\eqref{eq:attention_effect}), the spatial random effect from the \MBG{} framework \textit{$S(s_i, t_i)$} (Eq.~\eqref{eq:mbg_model}), and an independent nugget effect \textit{$\epsilon_i$}. This effectively models distinct types of variation: the continuous spatial field governed by physical proximity and a non-distance-based random effect captured by the attention process. This allows the model to learn structural dependencies, such as those driven by similar ecological features or conditions that do not strictly follow distance.

Assuming a binomial response, $Z_i \mid p_i,n_i \sim \mathrm{Binomial}(n_i,p_i)$, and $\operatorname{logit}(p_i) = \eta_i$, the proposed hybrid spatio-temporal model is given as:
\begin{equation}
    \eta_i =\beta_0 + \mathbf{x}_i^\top\boldsymbol{\beta} +\gamma \hat{O}_i +S(s_i,t_i) +A(s_i,t_i) +\epsilon_i,
\label{eq:spatio_temp_hybrid_model}
\end{equation}
The inclusion of the attention effect $A(s_i, t_i)$ enables the model to improve its precision by capturing structured residual dependencies that are not adequately represented by standard spatio-temporal random effects. 

This hybrid model is also applicable in the spatial domain. The resulting model 
\begin{equation}
    \eta_i =\beta_0 + \mathbf{x}_i^\top\boldsymbol{\beta} +\gamma \hat{O}_i +S(s_i) +A(s_i) +\epsilon_i,
\label{eq:spatial_hybrid_model}
\end{equation}
is the hybrid spatial model. 

\subsection{Models Comparison}\label{sec:models}
To identify which component of the hybrid is responsible for any gain, we fit nine models that differ only in the components they include (Table~\ref{tab:models}). Alongside the full hybrid model, we include: the hybrid with \emph{no} attention random effect; the hybrid with \emph{no} SPDE field; the hybrid with \emph{no} GATv2 offset; residual kriging after GATv2; and, in Section~\ref{sec:whatattn}, a graph effect built from a \emph{fixed} covariate-similarity graph rather than learned attention.

\begin{table}[htbp]
\centering
\caption{The nine fitted models. ``Offset'' is the GATv2 prediction
entered on the logit scale with an estimated coefficient; ``Attention''
is the \texttt{generic1} random effect built from the learned attention
Laplacian.}
\label{tab:models}
\small
\setlength{\tabcolsep}{3pt}
\begin{tabular}{lcccl}
\toprule
Model & Offset & SPDE & Attention & GATv2 graph edges \\
\midrule
\MBG{}            & --  & yes & --  & (no GATv2) \\
\Hyb{c}           & yes & yes & yes & covariate \\
\Hyb{tc}           & yes & yes & yes & time $+$ covariate \\
\Hyb{stc}         & yes & yes & yes & space $+$ time $+$ covariate \\
\Hyb{1x}          & yes & yes & yes & single covariate \\
\Hyb{w/o SPDE}    & yes & --  & yes & space $+$ time $+$ covariate \\
\Hyb{w/o Attn}    & yes & yes & --  & time $+$ covariate \\
\Hyb{w/o offset}  & --  & yes & yes & time $+$ covariate \\
GAT-KRI  & \multicolumn{4}{l}{GATv2 followed by spatio-temporal kriging of residuals} \\
\bottomrule
\end{tabular}
\end{table}

\subsection{Evaluation Metrics}
The performance of the competing models is assessed using the binomial log score, the denominator-weighted Brier score, empirical coverage, and calibration curves by risk group.

\subsubsection{Binomial Log Score}
The predictive performance of the model is measured using the binomial log score, by evaluating the log-likelihood of the observed binomial outcomes based on the estimated prevalence. The larger (less negative) the value, the better the prediction performance. The binomial log score is defined as
\begin{equation}
\text{LogScore} = \frac{1}{n} \sum_{i=1}^{n} \log\left[\binom{n_i}{z_i} \hat{p}_i^{\,z_i}
(1-\hat{p}_i)^{n_i-z_i}\right],
\end{equation}
where $n$ is the number of areas, $z_i$ is the observed number of cases in area $i$, $n_i$ is the number of individuals sampled (trials), $\hat{p}_i$ is the posterior mean estimate of the prevalence for area $i$.

\subsubsection{Denominator-Weighted Brier Score}
The denominator-weighted Brier score assesses the accuracy of the estimated prevalence by calculating the weighted mean squared error between the posterior mean estimates and the true prevalence values. The number of trials serves as a weight; thus, larger samples in an area have a greater influence on the overall measure. The smaller it is, the more accurate the estimate. The denominator-weighted Brier score is given by
\begin{equation}
\text{WBrier} = \frac{\sum_{i=1}^{n}
n_i\left(\hat{p}_i-p_i^{\text{true}}\right)^2}
{\sum_{i=1}^{n}n_i},
\end{equation}
where $n$ is the number of areas, $n_i$ is the number of trials in area $i$, $\hat{p}_i$ is the posterior mean prevalence estimate, $p_i^{\text{true}}$ is the true prevalence.

\subsubsection{Empirical Coverage}
Empirical coverage measures the proportion of true prevalence values that lie within the corresponding posterior credible intervals. It evaluates the calibration of the model's uncertainty estimates. A well-calibrated model should have empirical coverage close to the nominal coverage level (e.g., 0.95 for a 95\% credible interval). The empirical coverage is computed as
\begin{equation}
\text{Coverage}
=
\frac{1}{n}
\sum_{i=1}^{n}
I\left(L_i \le p_i^{\text{true}} \le U_i\right),
\end{equation}
where $n$ is the number of areas, $p_i^{\text{true}}$ is the true prevalence for area $i$, $L_i$ and $U_i$ are the lower and upper limits of the posterior credible interval, respectively, $I(\cdot)$ is the indicator function, which equals 1 if the condition is satisfied and 0 otherwise.

\subsubsection{Calibration Curve by Risk Group}
Calibration curves by risk group evaluate the agreement between predicted and true prevalence across various risk tiers. Areas are initially ordered according to their posterior-mean prevalence estimates and subsequently categorised into $G$ risk groups (e.g., deciles or quintiles). For every group, the average predicted prevalence is compared with the corresponding average true prevalence. A properly calibrated model generates points near the 45$^\circ$ line, indicating strong agreement between predicted and observed risk across the full range of prevalence. For the $g$-th risk group, the average predicted and true prevalence are computed as
\begin{equation}
\bar{\hat{p}}_g = \frac{1}{n_g}
\sum_{i \in g}\hat{p}_i, 
\quad and \quad 
\bar{p}^{\,\text{true}}_g = \frac{1}{n_g}
\sum_{i \in g}p_i^{\text{true}},
\end{equation}
where $G$ is the total number of risk groups, $n_g$ is the number of areas in risk group $g$, $\hat{p}_i$ is the posterior mean prevalence estimate, $p_i^{\text{true}}$ is the true prevalence for area $i$.

The calibration curve is obtained by plotting $\bar{\hat{p}}_g$ against $\bar{p}^{\,\text{true}}_g$ for $g=1,\ldots,G$, together with the identity line $y=x$. Closer agreement with the identity line indicates better calibration across the spectrum of predicted risk.

\section{Simulation Study}
\subsection{Design}
We assess the proposed hybrid model via Monte Carlo simulation across $9$ scenarios $\times$ $2$ hold-out schemes $\times$ $100$ replicates ($1800$ independent fits per model). Each replicate simulates $n=1200$ observations on $200$ spatial locations observed at $6$ time points. The scenarios are described in Table~\ref{tab:scenarios}. Writing $Z_i$ for the number of positives out of $n_i$ tested,
\begin{align}
Z_i \mid p_i, n_i &\sim \text{Binomial}(n_i, p_i), \label{eq:lik}\\
\operatorname{logit}(p_i) &= \beta_0 + \mathbf{x}_i^{\top}\boldsymbol{\beta}
 + f(\mathbf{x}_i) + S(s_i,t_i) + \psifield_i , \label{eq:eta}
\end{align}
with $n_i \sim \text{Uniform}\{30,\dots,80\}$, so that binomial denominators vary across units, as they do in real-life applications.

The four terms in \eqref{eq:eta} are switched on and off to define the scenarios: $\mathbf{x}_i^{\top}\boldsymbol{\beta}$ is a linear combination of ten covariates; $f(\cdot)$ is a nonlinear function of the same covariates; $S(s,t)$ is a distance-based Mat\'ern SPDE field with AR(1) temporal dependence ($\rho_t = 0.7$); and $\psifield$ is a non-distance \emph{similarity} field, drawn as a Gaussian Markov random field on a $k$-nearest-neighbour graph in covariate space. Both $S(s,t)$ and $\psifield$ are scaled to standard deviation $0.6$ on the logit scale, so no scenario is dominated by construction.

\begin{table}[htbp]
\caption{Number of edges and degree distribution}
\label{tab:deg_edges}
\small
\begin{tabular}{lrrrr}
\toprule
Edge set & Edges & Mean degree & Range & Isolated \\
\midrule
Covariate only            & $4\,503$  & $7.50$  & $5$--$19$ & $0$ \\
Covariate $+$ temporal    & $9\,103$  & $15.17$ & $7$--$32$ & $0$ \\
Covariate $+$ temporal $+$ spatial & $12\,549$ & $20.91$ & $11$--$41$ & $0$ \\
\bottomrule
\end{tabular}
\end{table}

The graph edges combine covariate-similarity edges based on all covariates, covariate-similarity edges based specifically on $X_{10}$, spatial edges, and temporal edges. $X_{10}$ was assigned the largest coefficient in the data-generating process and therefore had the strongest prespecified influence on the response. Table~\ref{tab:deg_edges} shows the number of edges and degree distribution. Degree exceeds $k$ because of symmetrisation: a node that is nobody else's neighbour has degree $5$, whereas a node in a dense region of covariate space accumulates more. The covariate graph has quartiles $(5, 6, 7, 9)$ with maximum $19$, and graph density $0.0063$. No node is isolated. After attention weighting, edge weights range from $0.0103$ to $0.2754$, with a median of $0.0623$.

\subsection{Validation Protocol}
Two hold-out schemes are used. Under random hold-out, $25\%$ of locations are withheld within each of five spatial strata, so a held-out site is unobserved at every time point; this is the interpolation task and is our primary result. Under \emph{block} hold-out, one spatial stratum is withheld together with a buffer, and the withheld block is rotated across replicates; this is a spatial-extrapolation stress test. All models are fitted to identical data and identical splits within a replicate.

\begin{table}[htbp]
\centering
\caption{Simulation scenarios. Each isolates one contrast; $\psifield$
is the non-distance similarity field.}
\label{tab:scenarios}
\footnotesize
\setlength{\tabcolsep}{4pt}
\begin{tabular}{lccll}
\toprule
& nonlinear $f$ & SPDE $S$ & $\psifield$ graph & purpose \\
\midrule
s1 & --  & yes & --                        & \MBG{} correctly specified \\
s2 & yes & yes & covariate                 & mixed dependence \\
s3 & yes & yes & covariate $+$ time        & mixed dependence \\
s4 & yes & yes & covariate $+$ time $+$ space & mixed dependence \\
s5 & yes & yes & single covariate          & sparse similarity \\
s6 & yes & --  & covariate $+$ time        & purely graph-based \\
s7 & yes & yes & --                        & no network effect \\
s8 & --  & yes & covariate $+$ time        & linear mean, graph dependence \\
s9 & --  & yes & \emph{misspecified}       & robustness to a wrong graph \\
\bottomrule
\end{tabular}
\end{table}

We construct an SPDE mesh with all nodes and standardise the features across the entire dataset. Similarly, the underlying graph of GATv2 spans all nodes. This transductive setting is usual for graph neural networks. Including the coordinates and covariates of held-out nodes during message delivery allows the model to acquire the spatial relationships needed to predict at those unseen locations. But for each replicate and hold-out split, we fit a linear pre-whitening regression of the empirical logit on the covariates using only the training rows, and use the fitted values to obtain residuals. Missing values are assigned to the residuals at all held-out nodes. These residuals are then used to train the GATv2 network. In this phase, the network learns the attention coefficients for all edges in the graph, effectively mapping the connectivity of the entire graph, but it learns them using the training data. The loss function and the internal validation split for early stopping are exclusively within the training set. Once the GATv2 network is optimised for the fold, we recompute the attention Laplacian $C$ for the full graph based on the fitted network weights. This Laplacian forms the structural basis for the last step: fitting the Bayesian model. This ensures that no response values from the held-out set are accessible at any point in the procedure until the final evaluation.

To determine the loss function to train the network, we compare MSE and Binomial NLL at production scale across 9 scenarios, with 100 replicates per scenario (Table~\ref{tab:loss_fxn_compared}). Within each replicate, the two losses are evaluated on the same data, train/test split, and random seed, giving a paired comparison that controls for replicate-level variation. Our main quantity of interest is the contribution of the attention random effect, measured by comparing the \Hyb{tc} to the \Hyb{w/o Attn}. The largest absolute difference between the two losses in the estimated attention effect is $0.53$ percentage points, whereas the attention effects themselves are approximately $21$--$27$ percentage points in the scenarios containing a similarity field (Table~\ref{tab:attn}). With 100 paired replicates, the comparison resolves differences of a few tenths of a percentage point, and at that resolution the two losses are distinguishable: four of the nine scenarios show a significant difference in the attention effect ($p$ between $0.0008$ and $0.030$), and in every scenario the Binomial NLL gives the slightly larger effect. The losses therefore agree closely but are not equivalent.

\begin{table}
\begin{center}
\small
\caption{Sensitivity of the attention effect to the choice of loss function. Values show the percentage change in test-set weighted Brier score for \Hyb{tc} relative to \Hyb{w/o Attn}, based on random hold-out evaluation across 100 replicates. Negative values indicate a lower weighted Brier score, favouring the attention effect.}
\label{tab:loss_fxn_compared}
\begin{tabular}{lrrrr}
\toprule
Scenario & MSE & Binomial NLL & Difference & 95\% CI \\
\midrule
s1 & $+0.5$  & $+0.4$  & $-0.10$ & $(-0.22,\ 0.02)$ \\
s2 & $-21.3$ & $-21.5$ & $-0.15$ & $(-0.44,\ 0.15)$ \\
s3 & $-22.3$ & $-22.6$ & $-0.30$ & $(-0.57,\ -0.04)$ \\
s4 & $-21.8$ & $-21.9$ & $-0.11$ & $(-0.37,\ 0.14)$ \\
s5 & $-21.3$ & $-21.8$ & $-0.53$ & $(-0.83,\ -0.22)$ \\
s6 & $-26.6$ & $-27.0$ & $-0.38$ & $(-0.82,\ 0.07)$ \\
s7 & $-7.4$  & $-7.8$  & $-0.35$ & $(-0.79,\ 0.10)$ \\
s8 & $-21.0$ & $-21.2$ & $-0.23$ & $(-0.43,\ -0.02)$ \\
s9 & $-20.5$ & $-20.8$ & $-0.33$ & $(-0.59,\ -0.07)$ \\
\bottomrule
\end{tabular}
\end{center}
\end{table}

We also compare the \Hyb{tc} against \MBG{}. Five of the nine scenarios show significant differences between the MSE-- and Binomial NLL--based estimates ($p$ between $0.003$ and $0.034$), with differences ranging from $-1.4$ to $-0.3$ percentage points and favouring the Binomial NLL in every scenario. These differences are small relative to the effects themselves and did not change the ordering of the scenarios or any qualitative conclusion. In the null scenario (s1), the \Hyb{tc} has a relative effect of $+4.9$\% under MSE and $+3.5$\% under Binomial NLL relative to \MBG{}. Coverage is essentially unchanged, with a maximum absolute difference of only $0.003$ across the nine scenarios. The choice of loss has little practical effect on the study's main conclusions. However, because the differences are not zero, we did not treat the two losses as interchangeable. We therefore fit the full study using the Binomial NLL.

\subsection{Prior Specification and Identifiability of Fitted Fields in \texorpdfstring{\Hyb{tc}}{Hyb(tc)}}
\label{sec:identif}

The model contains two structured latent effects, which could, in principle, represent overlapping structure: the SPDE field $S(s,t)$ encodes dependence that decays with geographic distance, while the attention effect $A_i$ encodes dependence along the graph. We therefore assess their separate identifiability directly rather than assuming it, in three ways: sensitivity of the variance components to the prior, the behaviour of the fitted fields themselves, and the behaviour of the model when the graph is deliberately
made distance-like.

\subsubsection{Sensitivity to the prior}
We vary the priors on the attention precision $\tau$ and the mixing parameter $\beta$ \emph{one at a time}, holding the SPDE priors fixed, so that any movement is attributable to a single hyperprior. Each setting is run across all nine scenarios with $100$ replicates and scored out-of-sample. Table~\ref{tab:prior_sensitivity_s3} reports scenario s3, in which the true similarity field has $\operatorname{sd}(\psi)=0.60$.

\begin{table}[htbp]
\centering
\caption{Prior sensitivity for the attention field, scenario s3, $100$ replicates. The priors on $\tau$ and $\beta$ are varied one at a time, with the SPDE priors held fixed throughout. $\tau^{-1/2}$ is the scale parameter of the attention field, not its marginal standard deviation: under Eq.~\eqref{eq:precision_matrix} the eigenvalues of $Q$ lie in $[\tau(1-\beta),\,\tau]$, so the marginal standard deviation of $A_i$ lies between $\tau^{-1/2}$ and $\tau^{-1/2}(1-\beta)^{-1/2}$. The prior column reports $\tau^{-1/2}$ evaluated at the prior mean of $\tau$. The true similarity field has $\operatorname{sd}(\psi)=0.60$.}
\label{tab:prior_sensitivity_s3}
\footnotesize
\setlength{\tabcolsep}{4pt}
\begin{tabular}{llrrrrr}
\hline
Prior on $\tau$ & Prior on $\beta$ & Prior $\tau^{-1/2}$ &
Pst. $\tau^{-1/2}$ & Pst. $\operatorname{sd}(S)$ &
Coverage & WBrier \\
\hline
$\mathrm{Gamma}(1,1)$        & $N(0,0.1)$ & $1.000$ & $0.506$ & $0.592$ & $0.937$ & $0.01882$ \\
$\mathrm{Gamma}(1,10^{-1})$  & $N(0,0.1)$ & $0.316$ & $0.476$ & $0.592$ & $0.935$ & $0.01881$ \\
$\mathrm{Gamma}(1,10^{-2})$  & $N(0,0.1)$ & $0.100$ & $0.474$ & $0.592$ & $0.935$ & $0.01881$ \\
\hline
$\mathrm{Gamma}(1,10^{-1})$  & $N(-2,1)$  & $0.316$ & $0.564$ & $0.596$ & $0.935$ & $0.01889$ \\
$\mathrm{Gamma}(1,10^{-1})$  & $N(0,1)$   & $0.316$ & $0.500$ & $0.592$ & $0.935$ & $0.01881$ \\
$\mathrm{Gamma}(1,10^{-1})$  & $N(2,1)$   & $0.316$ & $0.462$ & $0.591$ & $0.935$ & $0.01880$ \\
\hline
\end{tabular}
\end{table}

The posterior standard deviation of the SPDE field is unaffected by the prior on the attention effect: across two orders of magnitude of prior scale it varies by at most $0.005$ in s3, and if the two components were competing for the same residual structure, shrinking one would inflate the other. No such response is observed. The scale parameter of the attention field is likewise stable. As the prior mean of $\tau^{-1/2}$ is reduced from $1.000$ to $0.100$, a tenfold change, the posterior moves only from $0.506$ to $0.474$, and across all six prior settings it stays within $0.462$--$0.564$. The posterior therefore reflects the data rather than the prior. Because $\tau^{-1/2}$ bounds the marginal standard deviation of $A_i$ from below, these values are consistent with a fitted field of the scale of the true similarity component, whose standard deviation is $0.60$; the empirical standard deviation of $\hat A$ across the scenarios containing both structured effects is $0.48$--$0.53$.

Predictive conclusions are likewise insensitive. Across all six settings the weighted Brier score in s3 lies between $0.01880$ and $0.01889$ and coverage between $0.935$ and $0.937$, against $0.02431$ and $0.644$ for \MBG{}. This stability is a stronger form of the identifiability argument than the variance components alone provide: neither the magnitude of the attention effect nor the predictive gain it delivers can be manufactured by the choice of hyperprior. We adopt $\tau \sim \mathrm{Gamma}(1,1)$ and $\operatorname{logit}(\beta) \sim N(0,0.1)$ throughout, following the principle that a prior should be weakly informative on the scale of the effect it governs \citep{simpson2017penalising}.

\subsubsection{What each fitted field recovers}
Because the simulation knows the generating components, identifiability can be assessed more directly than through the variance components alone. We compare the posterior mean fields $\hat A$ and $\hat S$, evaluated at the data locations, against the two truths $\psi$ and $S$ (Table~\ref{tab:ident_fields}). These fits use the full data, since identifiability is a property of the fitted model rather than of its predictive behaviour.

\begin{table}[htbp]
\centering
\caption{Field-level identifiability diagnostics, $100$ replicates per scenario, Monte Carlo standard errors in brackets. $\psi$ is the similarity component of the generating process and $S$ the distance component.}
\label{tab:ident_fields}
\small
\begin{tabular}{lcccc}
\toprule
 & $\operatorname{cor}(\hat A,\psi)$ & $\operatorname{cor}(\hat A,S)$
 & $\operatorname{cor}(\hat S,S)$ & $\operatorname{cor}(\hat A,\hat S)$ \\
\midrule
s2 & $0.753$ $(0.003)$ & $0.151$ $(0.003)$ & $0.899$ $(0.003)$ & $0.153$ $(0.003)$ \\
s3 & $0.748$ $(0.003)$ & $0.145$ $(0.003)$ & $0.896$ $(0.004)$ & $0.149$ $(0.003)$ \\
s4 & $0.743$ $(0.003)$ & $0.149$ $(0.004)$ & $0.891$ $(0.003)$ & $0.160$ $(0.004)$ \\
s5 & $0.752$ $(0.003)$ & $0.146$ $(0.003)$ & $0.897$ $(0.003)$ & $0.157$ $(0.004)$ \\
s8 & $0.801$ $(0.002)$ & $0.147$ $(0.004)$ & $0.911$ $(0.002)$ & $0.144$ $(0.003)$ \\
s9 & $0.802$ $(0.001)$ & $0.149$ $(0.003)$ & $0.906$ $(0.002)$ & $0.152$ $(0.003)$ \\
\midrule
s1 (no $\psi$) & --- & $0.190$ $(0.003)$ & $0.958$ $(0.001)$ & $0.144$ $(0.003)$ \\
s7 (no $\psi$) & --- & $0.166$ $(0.003)$ & $0.945$ $(0.002)$ & $0.144$ $(0.003)$ \\
s6 (no $S$)    & $0.807$ $(0.003)$ & --- & --- & $0.289$ $(0.011)$ \\
\bottomrule
\end{tabular}
\end{table}

Each component recovers its own generating term. The attention field correlates $0.75$--$0.81$ with the similarity component and only $0.14$--$0.15$ with the distance component, while the SPDE field recovers
the distance component at $0.89$--$0.96$. In the six scenarios containing both structured effects the two fitted fields correlate at $0.145$--$0.161$; in s6, where there is no spatial field for the SPDE term to recover, the
correlation is $0.298$. Conditioning on $\hat A$ removes the SPDE field's association with $\psi$ almost entirely, with $\operatorname{cor}(\hat S,\psi \mid \hat A)$ between $-0.003$ and $0.019$ in those six scenarios ($0.064$ in s6), whereas $\operatorname{cor}(\hat A,\psi \mid \hat S)$ remains between $0.740$
and $0.799$. A standardised regression of $\psi$ on the three structured components attributes it to the attention field alone: the coefficient on $\hat A$ is $0.74$--$0.81$, against $-0.02$ to $0.05$ on $\hat S$ and
$0.03$--$0.08$ on the network offset. Removing the attention effect changes neither the recovery of $S$ nor its association with $\psi$ in any scenario containing both components. These correlations are invariant to rescaling
and are, as expected, unchanged under the prior settings of Table~\ref{tab:prior_sensitivity_s3}.

On the scale of the fields themselves, $\hat A$ has empirical standard deviation $0.48$--$0.53$ against $0.50$--$0.52$ for $\hat S$ across the six scenarios containing both effects, and accounts for $45$--$49\%$ of the
variance of the structured predictor; in s6, which contains no spatial field, it accounts for $93\%$.

Each effect also switches itself off when the corresponding structure is absent: where the generating process contains no similarity component, the attention field falls to a standard deviation of $0.08$ in s1 and $0.15$ in
s7, carrying $2\%$ and $7\%$ of the variance of the structured predictor against $45$--$49\%$ where a similarity field is present; and where there is no distance component, the SPDE field collapses instead to a standard deviation of $0.045$.

\subsubsection{Behaviour under a distance-aligned graph}
The diagnostics above come from scenarios in which the two truths are orthogonal by construction: the mean sample correlation between $\psi$ and $S$ is within $\pm 0.005$ in every scenario, and although individual
replicates reach $|r|\approx 0.11$, the average absolute correlation is only $0.023$. They therefore establish separability in the favourable case. To test the unfavourable one we refit the model varying
only the graph on which the attention effect is built, holding the data, the split and every other model term fixed. Three configurations are compared: the graph used throughout the paper, that graph with spatial
$k$-nearest-neighbour edges added, and a graph of spatial $k$-nearest neighbour edges alone.

\begin{table}[htbp]
\centering
\caption{Behaviour of the two structured effects when only the attention graph is changed, $100$ replicates per scenario. ``Default'' is the graph used throughout the paper, ``$+$sp'' adds spatial $k$-nearest-neighbour
edges, and ``sp only'' uses spatial $k$-nearest-neighbour edges alone. ``No $A$'' is the same model with the attention effect removed, and is unaffected by the graph.}
\label{tab:adv_graph}
\small
\setlength{\tabcolsep}{4pt}
\begin{tabular}{lcccc cccc c}
\toprule
& \multicolumn{4}{c}{$\operatorname{cor}(\hat S,S)$}
& \multicolumn{3}{c}{$\operatorname{sd}(\hat S)$}
& \multicolumn{2}{c}{$\operatorname{cor}(\hat A,\hat S)$} \\
\cmidrule(lr){2-5}\cmidrule(lr){6-8}\cmidrule(lr){9-10}
& Default & $+$sp & sp only & No $A$ & Default & $+$sp & sp only & Default & sp only \\
\midrule
s1 & $0.958$ & $0.954$ & $0.917$ & $0.961$ & $0.549$ & $0.476$ & $0.331$ & $0.144$ & $0.197$ \\
s2 & $0.899$ & $0.862$ & $0.816$ & $0.898$ & $0.503$ & $0.375$ & $0.293$ & $0.153$ & $0.187$ \\
s3 & $0.896$ & $0.857$ & $0.843$ & $0.894$ & $0.493$ & $0.347$ & $0.296$ & $0.149$ & $0.177$ \\
s4 & $0.891$ & $0.857$ & $0.815$ & $0.890$ & $0.501$ & $0.368$ & $0.286$ & $0.160$ & $0.201$ \\
s5 & $0.897$ & $0.866$ & $0.827$ & $0.897$ & $0.502$ & $0.384$ & $0.286$ & $0.158$ & $0.191$ \\
s8 & $0.911$ & $0.891$ & $0.844$ & $0.911$ & $0.516$ & $0.408$ & $0.301$ & $0.144$ & $0.178$ \\
s9 & $0.906$ & $0.883$ & $0.828$ & $0.906$ & $0.521$ & $0.409$ & $0.289$ & $0.152$ & $0.197$ \\
\bottomrule
\end{tabular}
\end{table}

A distance-aligned graph does induce competition between the two effects. Given a purely geographic graph, the attention effect draws variance out of the spatial field, with $\operatorname{sd}(\hat S)$ falling from about $0.51$ to $0.31$, and recovery of the distance component degrades to $0.817$--$0.848$ across the scenarios containing a similarity field. The same degradation appears in s1, where the generating process is purely distance-based and the attention effect has nothing it could legitimately explain: recovery falls
from $0.959$ to $0.917$. Under the graph used throughout the paper, none of this occurs: recovery of the distance component with the attention effect present matches its value in the model without the attention effect to three decimal places in six of the seven scenarios, and differs by at most $0.002$ in the seventh.

Identifiability of the two effects is therefore not automatic. It depends on the graph encoding associations that geographic proximity does not, which the construction used here does, and that a spatial adjacency graph would not. This is the same conclusion reached from the opposite direction by the ablation in Section~\ref{sec:results_ablation}, where performance improved as geographic edges were removed from the graph.

\subsection{Results}
\subsubsection{What the attention random effect contributes}
Table~\ref{tab:attn} isolates the attention effect by comparing \Hyb{tc} with \Hyb{w/o Attn}: the two models are identical except that the latter omits the attention random effect. The table shows that the contribution of the attention random effect depends strongly on whether a relational dependence exists. Where a similarity field $\psifield$ is present, the attention effect improves the test-set weighted Brier score by $21.2$--$27.0\%$ across all seven scenarios, with confidence intervals excluding zero and improvements observed in all $700$ replicates. The effect size is close to constant across those scenarios, as would be expected of a component capturing structure that the other terms cannot represent.

\begin{table}[htbp]
\centering
\caption{Contribution of the attention random effect: percentage change in test-set weighted Brier score from adding it to an otherwise identical model (\Hyb{tc} versus \Hyb{w/o Attn}), random hold-out, $100$ replicates. Negative values favour the attention effect.}
\label{tab:attn}
\small
\begin{tabular}{lcccc}
\toprule
Scenario & $\psifield$ present & \% change (95\% CI) & $p$ & replicates improved \\
\midrule
s1 & no  & $+2.8$ \;($+1.8$, $+3.8$)     & $4\times10^{-7}$ & 28/100 \\
s7 & no  & $-5.6$ \;($-6.7$, $-4.4$)     & $<10^{-14}$   & 83/100 \\
\midrule
s2 & yes & $-21.5$ ($-22.5$, $-20.5$)    & $<10^{-15}$   & \textbf{100/100} \\
s3 & yes & $-22.6$ ($-23.6$, $-21.5$)    & $<10^{-15}$   & \textbf{100/100} \\
s4 & yes & $-21.9$ ($-22.9$, $-20.9$)    & $<10^{-15}$   & \textbf{100/100} \\
s5 & yes & $-21.8$ ($-22.8$, $-20.8$)    & $<10^{-15}$   & \textbf{100/100} \\
s6 & yes & $-27.0$ ($-28.0$, $-26.0$)    & $<10^{-15}$   & \textbf{100/100} \\
s8 & yes & $-21.2$ ($-22.1$, $-20.4$)    & $<10^{-15}$   & \textbf{100/100} \\
s9 & yes & $-21.2$ ($-22.2$, $-20.2$)    & $<10^{-15}$   & \textbf{100/100} \\
\bottomrule
\end{tabular}
\end{table}

The two scenarios without a similarity field behave differently and are reported separately rather than averaged. In s7, which contains nonlinear covariate effects, the attention field still improves the Brier score by $5.6\%$: with no relational structure to capture, it instead absorbs part of the nonlinearity. In s1, which is linear and purely distance-based, there is nothing left for it to explain and including it is mildly harmful ($+2.8\%$, 95\% CI $[+1.8,+3.8]$, $p=4\times10^{-7}$). The cost is small in absolute terms, since the Brier score in s1 is an order of magnitude below that of the other scenarios, but it is not zero; a graph-structured effect carries some penalty where no relational dependence exists.

\subsubsection{Full ablation}\label{sec:results_ablation}
Table~\ref{tab:relative_scores} compares the predictive performance of all nine model specifications relative to \MBG{}. The results show a clear separation between the roles of the structured components. Removing the SPDE field is damaging only where distance-based dependence is the sole structure present: in s1 and s7, the two scenarios without a similarity field, \Hyb{w/o SPDE} is far worse than \MBG{} ($+282\%$ and $+113\%$ in weighted Brier). Wherever a similarity field is also present, the attention effect substitutes for much of what the SPDE would otherwise provide, and \Hyb{w/o SPDE} is close to \MBG{} or slightly better ($-4.9\%$ to $+2.8\%$ outside s6, and $-26.8\%$ in s6, the single scenario constructed without a spatial field). The two structured effects are therefore not redundant, but neither are they strictly complementary: the attention field can carry spatial signal when the graph contains geographically short edges, which is the mechanism examined directly in Section~\ref{sec:identif} and Table~\ref{tab:adv_graph}. Retaining both components is nonetheless the better specification throughout, since \Hyb{w/o SPDE} never matches the full hybrids in any scenario containing a spatial field.

\begin{table}[htbp]
\centering
\caption{Percentage change in test-set Binomial Log Score and Weighted Brier score relative to
\MBG{}, random hold-out, $100$ replicates. Negative values are better
than \MBG{}.}
\label{tab:relative_scores}
\footnotesize
\setlength{\tabcolsep}{4.5pt}
\begin{tabular}{lcccccccc}
\toprule
& \Hyb{c} & \Hyb{tc} & \Hyb{stc} & \Hyb{1x} &
\Hyb{w/o SPDE} & \Hyb{w/o Attn} & \Hyb{w/o off} & GAT-KRI \\
\midrule

\multicolumn{9}{l}{\textit{Binomial Log Score}} \\
s1 & $+0.2$ & $+0.5$ & $+1.3$ & $+0.3$ & $+25.0$ & $+0.3$ & $+0.1$ & $+55.7$ \\
s2 & $-13.1$ & $-12.4$ & $-11.3$ & $-12.3$ & $-3.1$ & $+0.4$ & $-13.2$ & $+12.4$ \\
s3 & $-13.6$ & $-12.9$ & $-11.3$ & $-12.5$ & $-3.7$ & $+0.1$ & $-13.6$ & $+11.5$ \\
s4 & $-12.9$ & $-12.1$ & $-11.0$ & $-11.8$ & $-3.1$ & $+0.4$ & $-12.8$ & $+13.6$ \\
s5 & $-12.7$ & $-12.0$ & $-10.6$ & $-11.9$ & $-2.5$ & $+0.5$ & $-12.7$ & $+13.4$ \\
s6 & $-15.3$ & $-14.9$ & $-14.6$ & $-14.0$ & $-14.6$ & $-0.1$ & $-15.3$ & $+6.8$ \\
s7 & $-2.6$ & $-1.4$ & $-0.3$ & $-1.4$ & $+18.4$ & $-0.0$ & $-1.9$ & $+44.7$ \\
s8 & $-11.3$ & $-11.0$ & $-9.2$ & $-10.8$ & $+0.3$ & $+0.4$ & $-11.5$ & $+18.9$ \\
s9 & $-11.3$ & $-10.9$ & $-9.2$ & $-10.7$ & $+0.7$ & $+0.5$ & $-11.4$ & $+18.3$ \\

\midrule
\multicolumn{9}{l}{\textit{Weighted Brier}} \\
s1 & $+2.9$ & $+6.0$ & $+15.2$ & $+3.8$ & $+281.7$ & $+3.1$ & $+1.7$ & $+626.6$ \\
s2 & $-22.4$ & $-21.1$ & $-19.1$ & $-20.9$ & $-3.6$ & $+0.5$ & $-22.6$ & $+25.3$ \\
s3 & $-23.5$ & $-22.2$ & $-19.3$ & $-21.5$ & $-4.9$ & $+0.4$ & $-23.6$ & $+23.1$ \\
s4 & $-22.7$ & $-21.4$ & $-19.2$ & $-20.8$ & $-3.9$ & $+0.6$ & $-22.7$ & $+26.9$ \\
s5 & $-22.3$ & $-21.0$ & $-18.6$ & $-20.9$ & $-3.6$ & $+1.0$ & $-22.4$ & $+25.3$ \\
s6 & $-27.8$ & $-27.3$ & $-26.6$ & $-25.4$ & $-26.8$ & $-0.3$ & $-28.1$ & $+14.4$ \\
s7 & $-10.2$ & $-5.3$ & $+0.7$ & $-5.3$ & $+113.0$ & $+0.3$ & $-7.8$ & $+259.6$ \\
s8 & $-21.3$ & $-20.6$ & $-17.1$ & $-20.4$ & $+1.6$ & $+0.8$ & $-21.7$ & $+38.1$ \\
s9 & $-21.2$ & $-20.5$ & $-17.3$ & $-20.2$ & $+2.8$ & $+0.9$ & $-21.5$ & $+36.9$ \\

\bottomrule
\end{tabular}
\end{table}

Removing the attention effect changes performance only modestly across most scenarios, with differences generally within a few percentage points, whereas its inclusion produces the larger gains seen in the table when a similarity field exists. This suggests that the attention component contributes information that is complementary to, rather than a substitute for, the dominant spatial structure. Residual kriging after GATv2 is worse
than \MBG{} in every scenario. Therefore, it does not recover the structured dependence as effectively as the joint model specification.

Two aspects of the ablation bear on the choice of graph for the attention effect. All comparisons below are paired within replicate and pooled over the seven scenarios containing a similarity field, under the random hold-out reported in the table. First, removing the GATv2 offset does not degrade performance: \Hyb{w/o offset} improves on \Hyb{tc} by $1.47\%$ (95\% CI $1.24$--$1.70$, $p<10^{-34}$). The nonlinear structure that the offset is intended to carry is already represented by the attention field and the covariate terms. Second, the composition of the graph matters more than its size. \Hyb{c}, whose graph uses covariate-similarity edges alone, improves on \Hyb{tc} by $1.23\%$ ($p<10^{-22}$), while \Hyb{stc}, which adds spatial $k$-nearest-neighbour edges, is worse by $3.14\%$ ($p<10^{-51}$). The same ordering holds under the
block hold-out, where the differences are larger ($1.72\%$, $1.43\%$ and $5.39\%$ respectively when both schemes are pooled).

These results admit a common explanation. The temporal edges of the graph are constructed by geographic nearest-neighbour search across time slices and are consequently very short, whereas the covariate-similarity edges are
geographically unstructured. Performance improves as the geographic component of the graph is removed and degrades as it is added. This is consistent with the framework, in which the attention field is useful as far as it encodes association that the distance-based field cannot.

Table~\ref{tab:coverage} reports empirical coverage of nominal $95\%$ credible intervals for the latent risk $p_i$. \MBG{} under-covers severely wherever a similarity field is present, reaching $0.499$ in s6: with no mechanism to represent non-distance dependence, it is substantially overconfident. Adding the attention effect raises coverage by $27$ to $44$ percentage points, to $0.93$--$0.94$, and does so consistently: across the seven scenarios containing a similarity field, every hybrid that carries both structured effects lies between $0.929$ and $0.942$, a narrow band a little below the nominal $0.95$. The improvement is attributable to the attention effect specifically: \Hyb{w/o Attn} is indistinguishable from \MBG{} throughout. In the two scenarios without a similarity field, those same models cover at $0.96$--$0.98$, slightly above nominal, reflecting the mild conservatism noted in Section~\ref{sec:identif}.

\begin{table}[htbp]
\centering
\caption{Empirical coverage of nominal $95\%$ credible intervals for the latent risk $p_i$, test set, random hold-out, $100$ replicates.}
\label{tab:coverage}
\footnotesize
\begin{tabular}{lrrrrrrrrr}
\toprule
& s1 & s2 & s3 & s4 & s5 & s6 & s7 & s8 & s9 \\
\midrule
\MBG{}           & $0.919$ & $0.649$ & $0.644$ & $0.657$ & $0.649$ & $0.499$ & $0.884$ & $0.656$ & $0.663$ \\
\Hyb{c}          & $0.979$ & $0.934$ & $0.940$ & $0.939$ & $0.939$ & $0.938$ & $0.966$ & $0.936$ & $0.941$ \\
\Hyb{tc}         & $0.978$ & $0.933$ & $0.937$ & $0.938$ & $0.937$ & $0.937$ & $0.962$ & $0.932$ & $0.939$ \\
\Hyb{stc}        & $0.974$ & $0.929$ & $0.933$ & $0.936$ & $0.934$ & $0.935$ & $0.957$ & $0.929$ & $0.935$ \\
\Hyb{1x}         & $0.979$ & $0.933$ & $0.936$ & $0.937$ & $0.937$ & $0.933$ & $0.962$ & $0.932$ & $0.937$ \\
\Hyb{w/o SPDE}   & $0.955$ & $0.933$ & $0.937$ & $0.939$ & $0.936$ & $0.935$ & $0.952$ & $0.934$ & $0.936$ \\
\Hyb{w/o Attn}   & $0.917$ & $0.655$ & $0.654$ & $0.667$ & $0.655$ & $0.523$ & $0.885$ & $0.658$ & $0.669$ \\
\Hyb{w/o offset} & $0.980$ & $0.936$ & $0.940$ & $0.942$ & $0.940$ & $0.939$ & $0.967$ & $0.937$ & $0.940$ \\
GAT-KRI & $0.891$ & $0.925$ & $0.923$ & $0.927$ & $0.928$ & $0.938$ & $0.900$ & $0.913$ & $0.916$ \\
\bottomrule
\end{tabular}
\end{table}

The results for the block hold-out scheme are reported in ~\ref{app:supplementary}. The overall pattern is similar to that observed under random hold-out. The block hold-out results provide a consistent spatial-extrapolation stress test; however, it does not substantially alter the conclusions drawn from the random hold-out analysis.

\begin{figure}[htbp]
\centering
\includegraphics[width=0.9\textwidth]{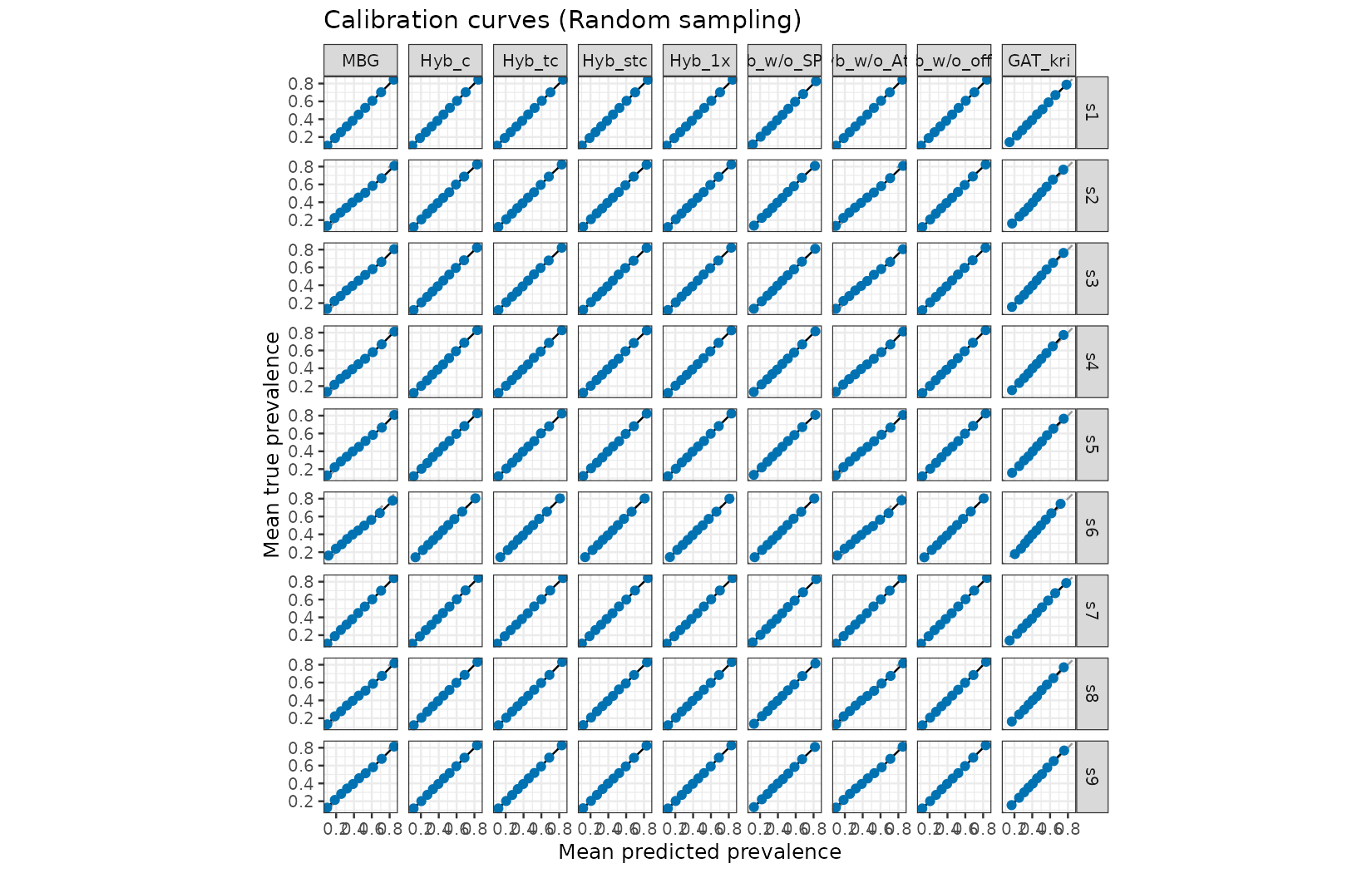}
\caption{Calibration by decile of predicted risk, test set, random hold-out, pooled over $100$ replicates. Rows are scenarios, columns are models; the dashed line is perfect calibration.}
\label{fig:calib}
\end{figure}

\subsubsection{Is it the graph, or merely an extra random effect?}\label{sec:whatattn}
A structured graph effect could, in principle, be mimicked by any node-level random effect with a well-chosen prior. To test this, we fit four full hybrid models that are identical except for the node-level term: the learned attention Laplacian; a \emph{fixed} covariate $k$-NN Laplacian with binary edge weights and no learning; an unstructured i.i.d. effect with two specifications of precision prior; and no node-level term at all. Each is run over all nine scenarios with $100$ replicates (Table~\ref{tab:whatattn}). The fixed comparator is built from within-time covariate similarity alone, so it differs from the learned Laplacian in two respects: binary rather than learned edge weights, and the absence of the temporal edge block.

The graph structure, rather than the addition of a random effect, is what produces the improvement, and the effect is large. Replacing the graph-structured term with an unstructured i.i.d.\ effect of the same default prior $Gamma(1,1)$ costs $21$--$27\%$ in Brier score in every scenario containing a similarity field, in $100$ of $100$ replicates ($p<10^{-60}$ throughout), and its coverage ($0.52$--$0.67$) is as poor as \MBG{}'s. This is expected: an unstructured effect carries no information to a held-out node, whereas a graph-structured effect borrows strength from that node's observed neighbours. The i.i.d.\ effect is in fact slightly worse than omitting the node-level term altogether, and markedly so in the two scenarios without a similarity field ($+10.1\%$ and $+5.1\%$, both $p<10^{-27}$). 

\begin{table}[htbp]
\centering
\caption{Four models differing in the node-level random effect and prior precision parameter $\tau$, $\mathcal{P}_1$ denotes $\tau \sim Gamma(1,1)$ and $\mathcal{P}_2$ denotes $\tau \sim Gamma(1,10^{-1})$, with $100$ replicates per scenario. 
The final column is the paired percentage difference between the learned attention Laplacian and the fixed covariate graph; $^{*}p<0.05$, $^{***}p<0.001$.}
\label{tab:whatattn}
\small
\begin{tabular}{lrrrrrrr}
\toprule
& \multicolumn{4}{c}{WBrier} & \multicolumn{2}{c}{Coverage} & learned vs \\
\cmidrule(lr){2-5}\cmidrule(lr){6-7}
 & attention & fixed & i.i.d. & none & attention & fixed & fixed (\%) \\
\midrule
\multicolumn{7}{l}{\textit{Prior $\mathcal{P}_1$}} \\
s1 & $0.0023$ & $0.0023$ & $0.0024$ & $0.0022$ & $0.978$ & $0.977$ & $+0.08$ \\
s2 & $0.0192$ & $0.0192$ & $0.0245$ & $0.0245$ & $0.932$ & $0.933$ & $+0.18^{**}$ \\
s3 & $0.0189$ & $0.0188$ & $0.0245$ & $0.0245$ & $0.937$ & $0.939$ & $+0.79^{***}$ \\
s4 & $0.0186$ & $0.0185$ & $0.0239$ & $0.0238$ & $0.937$ & $0.938$ & $+0.38^{***}$ \\
s5 & $0.0194$ & $0.0194$ & $0.0248$ & $0.0247$ & $0.935$ & $0.936$ & $+0.11$ \\
s6 & $0.0163$ & $0.0162$ & $0.0224$ & $0.0223$ & $0.937$ & $0.940$ & $+0.56^{***}$ \\
s7 & $0.0047$ & $0.0047$ & $0.0053$ & $0.0050$ & $0.962$ & $0.963$ & $+0.29$ \\
s8 & $0.0167$ & $0.0166$ & $0.0212$ & $0.0211$ & $0.933$ & $0.935$ & $+0.14$ \\
s9 & $0.0167$ & $0.0167$ & $0.0212$ & $0.0211$ & $0.938$ & $0.938$ & $-0.01$ \\
\midrule
\multicolumn{7}{l}{\textit{Prior $\mathcal{P}_2$}} \\
s1 & $0.0023$ & $0.0023$ & $0.0023$ & $0.0022$ & $0.954$ & $0.955$ & $+0.06$ \\
s2 & $0.0192$ & $0.0192$ & $0.0245$ & $0.0245$ & $0.931$ & $0.930$ & $+0.15$ \\
s3 & $0.0189$ & $0.0188$ & $0.0245$ & $0.0245$ & $0.935$ & $0.935$ & $+0.69^{***}$ \\
s4 & $0.0186$ & $0.0185$ & $0.0239$ & $0.0238$ & $0.935$ & $0.936$ & $+0.34^{***}$ \\
s5 & $0.0194$ & $0.0193$ & $0.0247$ & $0.0247$ & $0.931$ & $0.934$ & $+0.34^{*}$ \\
s6 & $0.0163$ & $0.0162$ & $0.0223$ & $0.0223$ & $0.935$ & $0.937$ & $+0.47^{***}$ \\
s7 & $0.0046$ & $0.0045$ & $0.0051$ & $0.0050$ & $0.953$ & $0.953$ & $+1.53^{***}$ \\
s8 & $0.0167$ & $0.0167$ & $0.0211$ & $0.0211$ & $0.931$ & $0.931$ & $+0.03$ \\
s9 & $0.0168$ & $0.0167$ & $0.0212$ & $0.0211$ & $0.932$ & $0.936$ & $+0.58^{*}$ \\
\bottomrule
\end{tabular}
\end{table}

The learned attention Laplacian does not, however, outperform the fixed comparator. Under $\mathcal{P}_1$ the fixed graph is ahead by $0.28\%$ pooled across the nine scenarios ($95\%$ CI $+0.21$ to $+0.35$, $n=900$), reaching significance in four scenarios, and under $\mathcal{P}_2$ it is ahead by $0.47\%$ ($95\%$ CI $+0.36$ to $+0.58$, $n=900$), reaching significance in six. The ordering is therefore the same under both priors, and in no scenario does the learned graph lead by a significant margin. The differences are small in absolute terms, never exceeding $1.6\%$ and typically below $0.8\%$, which is an order of magnitude less than the $21$--$27\%$ cost of removing the graph structure and smaller than the $3.1\%$ penalty incurred by adding spatial edges.

This comparison should not be read as establishing that binary weights are preferable, because it does not isolate the weighting. The fixed comparator is built from within-time covariate similarity, so it differs from the learned Laplacian in two respects at once: binary rather than learned edge weights, and the absence of the temporal edge block. Section~\ref{sec:results_ablation} shows that removing the temporal edges is independently worth about $1.2\%$, so a difference of the size observed is equally consistent with a learned weighting of that order being offset by the edges it carries. Separating the two would require a comparison holding topology fixed while varying only the weights, which we did not run.

What the experiment does establish is the distinction the comparison was designed to test. A graph-structured node effect is necessary, and its contribution is large; an unstructured effect of the same prior recovers none of it, and the composition of the graph matters, with non-Euclidean edges improving performance and geographic edges reducing it. The evidence does not support attributing the gain specifically to the learned edge weights, and the claims made in this paper are set accordingly.

\subsection{What the simulation does and does not establish}\label{sec:limits}
Two questions should be distinguished here: whether the attention field recovers the similarity structure, and whether the predictive gain depends on that recovery. The evidence differs between the two, and both are
reported.

The fitted attention field does recover the similarity component. As shown in Section~\ref{sec:identif}, $\operatorname{cor}(\hat A,\psi)$ is $0.74$--$0.81$ across the scenarios containing one; the association survives conditioning on the spatial field, and a standardised regression of $\psi$ on the structured components of the linear predictor attributes it to $\hat A$ alone.

The predictive gain is tied to that recovery. Holding graph dimension fixed at two covariates and varying only whether those covariates are the ones generating the similarity field, the correctly specified graph reduces the weighted Brier score by $34.3\%$ relative to the misspecified one ($p<10^{-3}$, $20$ replicates; mean Brier $0.0084$ against $0.0129$). The content of the graph, and not merely its dimension, therefore determines how much the attention effect contributes. This is the comparison that isolates the question, because the two graphs are identical in size and differ only in which covariates define similarity.

We therefore claim both that the attention field represents structured non-Euclidean association, recovering the generating similarity component in these simulations, and that the predictive improvement depends on the graph encoding the relevant similarity rather than an arbitrary one. What the experiments do not establish is that the \emph{learned weighting} of a given edge set is what delivers the gain, as Section~\ref{sec:whatattn} sets out.

\section{Application to Malaria Prevalence Data in Nigeria}
\subsection{Data}
Malaria remains one of the most devastating parasitic diseases affecting humans, presenting a significant global health challenge. Despite being both preventable and treatable, the disease continues to exert a heavy toll on public health systems, particularly in Africa. In 2022, the continent accounted for approximately 94\% of all malaria cases (233 million) and 95\% of malaria deaths (580,000) \citep{WHO2023WorldMalaria}. This disparity is largely driven by the dominance of \textit{Plasmodium falciparum}, the most lethal malaria parasite species \citep{snow2017prevalence}. Nigeria bears the heaviest burden, accounting for approximately 27\% of the global malaria burden and 31\% of global malaria deaths, making it the country with the highest number of cases and fatalities worldwide (WHO, 2023). Malaria is transmitted to humans through the bite of infected female \textit{Anopheles} mosquitoes. In Nigeria and broader West Africa, the primary vectors belong to the \textit{Anopheles gambiae} complex and  \textit{Anopheles funestus} group \citep{sinka2010dominant}.

This disease is a useful setting for a relational modelling approach because transmission is influenced by processes that are not captured by geographic proximity alone. Environmental conditions affect mosquito abundance, parasite development, and human exposure, while human movement and shared ecological conditions can create connections between populations that are geographically separated. As a result, two locations may have similar malaria prevalence because they share similar characteristics, even when they are far apart. Similarly, nearby locations may have quite different levels of malaria when their environmental conditions or local transmission contexts differ. A conventional spatial model does not directly account for these types of relationships because its dependence is primarily defined by geographic proximity.

In the hybrid model, the graph provides a way to represent these relationships using the characteristics of the observations. The attention mechanism then learns how strongly these related observations should influence one another. This application is therefore more than a demonstration of predictive performance. It provides a setting in which the relational component has a meaningful interpretation and can be assessed alongside conventional spatial dependence.

\begin{figure}[t]
    \centering
    \includegraphics[width=1\linewidth]{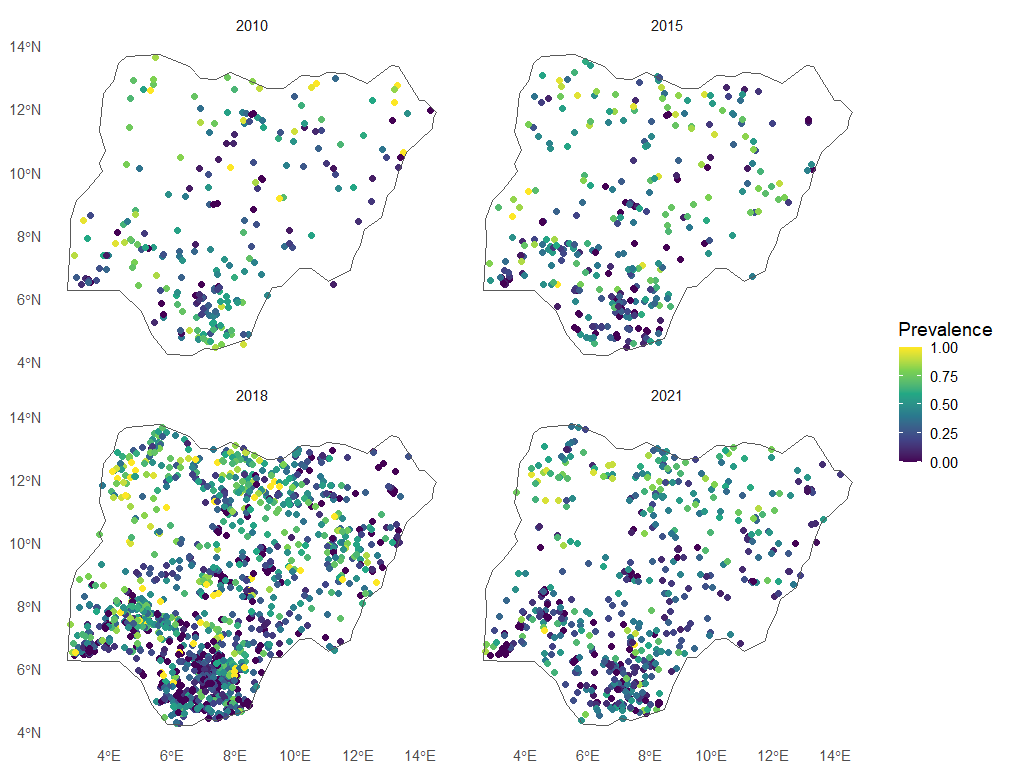}
    \caption{Map of Nigeria showing observed malaria prevalence in the years 2010, 2015, 2018, and 2021}
    \label{fig:Obs Prevalence}
\end{figure}

A complex array of environmental covariates drives the high prevalence in Nigeria. The following covariates were used for this study: Enhanced Vegetation Index (EVI), Normalised Difference Vegetation Index(NDVI), distance to water, elevation, potential evaporation, relative humidity, skin temperature, slope, soil temperature, surface runoff, sub-surface runoff, surface net solar radiation, temperature, max temperature, min temperature, total evaporation, total precipitation and volumetric soil water,
For definitions and resolutions, see Table~\ref{tab:def. of covs}.

\subsection{Data Source}
The study used a survey dataset from the Demographic and Health Surveys (DHS) for Nigeria \citep{NigeriaDHS_Grouped}. Malaria rapid diagnostic test (RDT) results are available only for the years 2010, 2015, 2018, and 2021 survey phases. Individual-level malaria RDT outcomes (positive or negative) were obtained from the household-member recode files, along with the corresponding cluster identifiers. For each survey cluster, the total number of individuals tested and the number of positive RDT results were aggregated to construct cluster-level malaria outcomes. These cluster-level outcomes were then linked to the DHS cluster geographic coordinates using the cluster identifiers. Given that DHS cluster coordinates are randomly displaced for confidentiality by up to 2 km in urban areas and 5 km in rural areas, with approximately 1\% of rural clusters displaced by up to 10 km  \citep{NigeriaDHS_Grouped}, environmental covariates were extracted at cluster locations and other unobserved locations at 10--km spatial resolution using Google Earth Engine \citep{GORELICK201718} \url{https://earthengine.google.com/}, to reduce the potential influence of coordinate displacement on covariate assignment.

Climate variables were sourced from the ERA5--Land Monthly Aggregated dataset \citep{sabater2019era5}. To derive annual values, monthly mean variables, including the temperature covariates and volumetric soil water, were averaged, while monthly accumulated variables, including total precipitation, runoffs, evaporation, and surface net solar radiation, were summed. Relative humidity was calculated from monthly air and dew-point temperatures and subsequently averaged for the year. Vegetation indices (NDVI and EVI) were obtained from the MODIS MOD13A3 product \citep{didan2021modis} and averaged annually. Topographic data (elevation and slope) and distance to water were derived from the 30m Shuttle Radar Topography Mission (SRTM) \citep{farr2007srtm} and Joint Research Centre (JRC) Global Surface Water datasets \citep{pekel2016high}, respectively. All covariates were aligned with the specific year and coordinates of each survey observation.

The malaria data from the four survey years comprise $2,480$ clusters and $32,653$ individuals with valid RDT results. The numbers of clusters, individuals tested, and positive RDT results by survey year are reported in Table~\ref{tab:survey_summary}. The number of individuals tested per cluster ranged from $1$ to $84$, with a mean of $13.17$ (SD $9.17$). Individuals with missing RDT results were excluded from the analysis.

\begin{table}[htbp]
\centering
\caption{Number of clusters, individuals tested, and positive RDT results by survey year.}
\footnotesize
\setlength{\tabcolsep}{5pt}
\label{tab:survey_summary}
\begin{tabular}{lrrrr}
\toprule
Year & Observed clusters & Individuals tested & Positive RDTs & Prediction-grid locations \\
\midrule
2010 & 237  & 5,122  & 2,412 & 9490 \\
2015 & 320  & 5,862  & 2,547 & 9490 \\
2018 & 1,366 & 11,071 & 4,226 & 9490 \\
2021 & 557  & 10,598 & 4,068 & 9490 \\
\midrule
Total & 2,480 & 32,653 & 13,253 & 37,960 \\
\bottomrule
\end{tabular}
\end{table}

\subsection{Covariate Selection}
To obtain a robust \MBG{} model for comparison with the proposed hybrid approach, we implemented a multi-stage selection procedure. First, we used all the covariates as a specification. Secondly, we assessed multicollinearity among the covariates by calculating the Variance Inflation Factor (VIF) for each covariate. Covariates exhibiting high multicollinearity ($VIF\geq5$) were iteratively removed through stepwise elimination. The remaining covariates were then subjected to Regularised Variable Selection using LASSO regression \citep{tibshirani1996regression}. Concurrently, we conducted an exhaustive search by fitting geostatistical \MBG{} models for all possible subsets of the non-collinear covariates. These models were ranked according to the Watanabe–Akaike Information Criterion (WAIC) and the Deviance Information Criterion (DIC), with lower values indicating better model fit, and by the Log Mean Conditional Predictive Ordinate (LCPO), with higher values indicating superior predictive performance. 

\subsection{Results}
Following the model selection procedure, LASSO regression identified a covariate subset comprising EVI, distance to water, elevation, potential evaporation, slope, subsurface runoff, surface runoff, total evaporation, and volumetric soil water. The WAIC and DIC metrics from the exhaustive model search selected a model with covariates EVI, slope, subsurface runoff, surface runoff, and volumetric soil water, and the log mean CPO metric selected a model containing EVI and elevation. We identified four candidate covariate specifications for further evaluation: the complete set of covariates (\textsc{all-cov}), the subset selected via LASSO regression (\textsc{lasso}), the subset chosen by WAIC and DIC (\textsc{ic}), and the subset selected using the Log Mean Conditional Predictive Ordinate (\textsc{cpo}). Because EVI was retained in all candidate specifications, it was the common covariate across the model-selection approaches and was therefore selected to define the covariate-similarity edges for the \Hyb{1x} model.

\begin{figure}[H]
        \centering
        \includegraphics[width=1\linewidth]{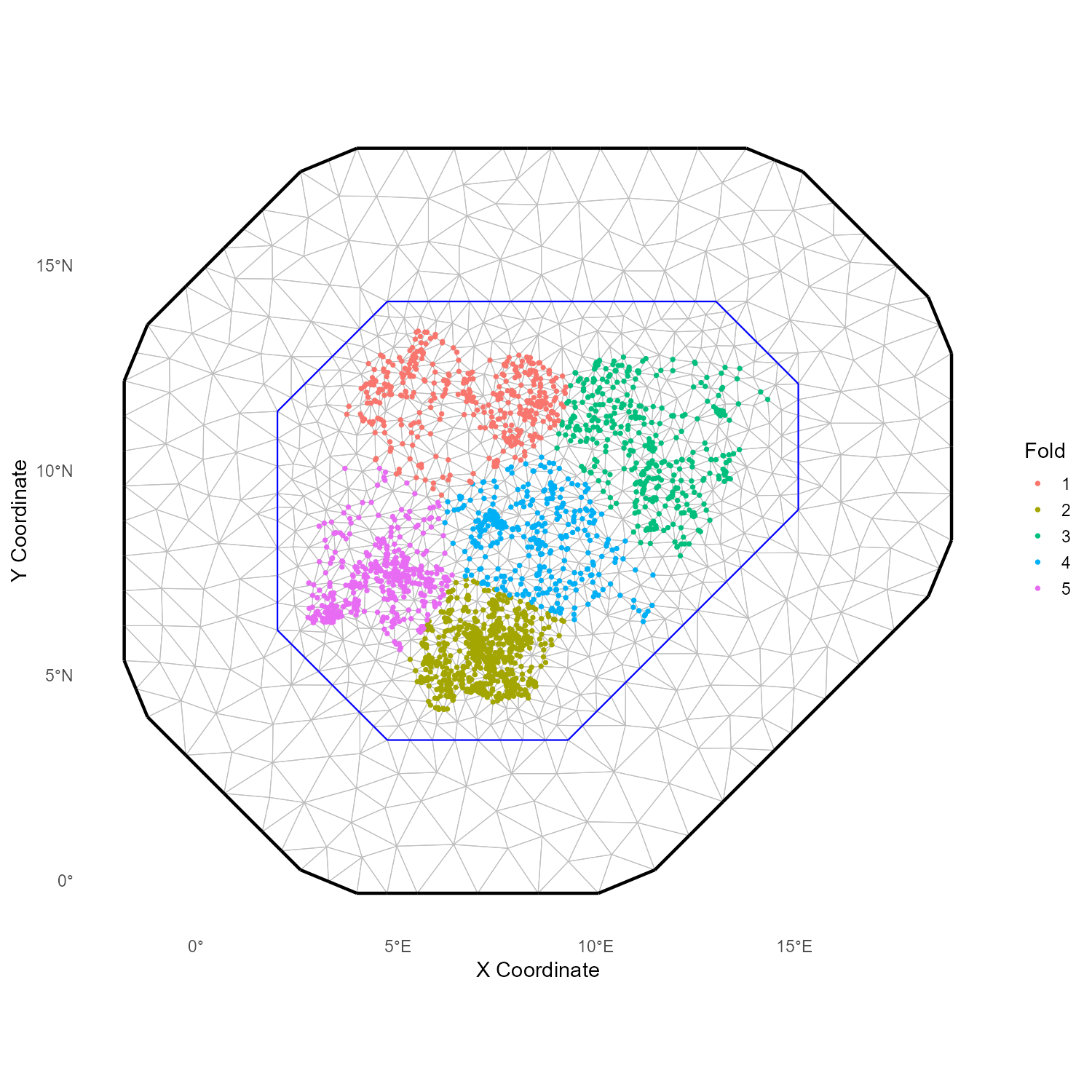}
        \caption{Spatial clusters for cross-validation over the spatial mesh}
        \label{fig:Spatial Clusters}
\end{figure}

We partition all the locations into 5 folds for spatial cross-validation (Fig.\ref{fig:Spatial Clusters}). In each iteration, four folds were used to train the models and the remaining fold was withheld for validation, with each fold serving as the validation set once. Each specification of environmental covariates was standardised within each iteration using the mean and standard deviation estimated from the training data in each fold, and the same transformation was subsequently applied to the corresponding test data to prevent information leakage. The predictions were subsequently obtained for the held-out clusters, and performance metrics were estimated and compared across models.

\begin{table}[htbp]
\centering
\caption{Percentage change in test-set Binomial Log Score and Weighted Brier score relative to \MBG{} across alternative covariate specifications. Negative
values indicate improvement over \MBG{}.}
\label{tab:covariate_comparison}
\footnotesize
\setlength{\tabcolsep}{2.5pt}
\begin{tabular}{lcccccccc}
\toprule
& \multicolumn{2}{c}{\textsc{all-cov}}
& \multicolumn{2}{c}{\textsc{cpo}}
& \multicolumn{2}{c}{\textsc{ic}}
& \multicolumn{2}{c}{\textsc{lasso}} \\
\cmidrule(lr){2-3}
\cmidrule(lr){4-5}
\cmidrule(lr){6-7}
\cmidrule(lr){8-9}
& LogScore & WBrier & LogScore & WBrier & LogScore & WBrier & LogScore & WBrier \\
\midrule
\Hyb{c} & $-6.4$ & $-6.9$ & $-17.6$ & $-21.2$ & $-14.6$ & $-17.5$ & $-17.5$ & $-22.1$ \\
\Hyb{tc} & $-5.6$ & $-6.0$ & $-17.1$ & $-20.4$ & $-14.5$ & $-17.4$ & $-14.6$ & $-18.4$ \\
\Hyb{stc} & $-2.6$ & $-2.2$ & $-21.4$ & $-26.6$ & $-18.3$ & $-22.9$ & $-14.8$ & $-18.4$ \\
\Hyb{1x} & $-7.7$ & $-8.4$ & $-14.4$ & $-16.6$ & $-14.3$ & $-17.0$ & $-14.0$ & $-16.8$ \\
\Hyb{w/o SPDE} & $+0.4$ & $+2.0$ & $-20.4$ & $-25.4$ & $-19.4$ & $-24.6$ & $-13.7$ & $-17.1$ \\
\Hyb{w/o Attn} & $-2.4$ & $-3.1$ & $-5.5$ & $-6.9$ & $+1.6$ & $+2.2$ & $+0.6$ & $+1.3$ \\
\Hyb{w/o offset} & $-7.7$ & $-8.2$ & $-14.9$ & $-17.2$ & $-14.6$ & $-17.3$ & $-15.1$ & $-18.4$ \\
GAT-KRI & $+33.6$ & $+38.8$ & $+14.9$ & $+6.7$ & $+16.9$ & $+11.3$ & $+17.1$ & $+12.1$ \\
\bottomrule
\end{tabular}
\end{table}

Table~\ref{tab:covariate_comparison} investigates the robustness of the hybrid model to alternative covariate specifications. An important point of interpretation is that covariate selection strategies also affect the \MBG{} baseline. Relative to the \MBG{} model fitted with all covariates, the \MBG{} fitted with CPO-selected covariates has a $21.3\%$ lower LogScore and a $38.3\%$ higher WBrier score. The corresponding changes are $16.6\%$ and $31.3\%$ for IC, and $14.9\%$ and $27.1\%$ for LASSO. Thus, the selected-covariate specifications for \MBG{} models are less accurate than the \MBG{} model with all covariates. The hybrid results should therefore be read as improvements relative to each matched baseline rather than as evidence that one covariate specification produces the best absolute predictive performance based on the results in the table alone.

Against these matched baselines, comparing \Hyb{tc} with \Hyb{w/o\_Attn} gives a direct assessment of the contribution of the attention effect in this malaria study. The size of that contribution depends strongly on how rich the covariate specification is. Under the three parsimonious specifications, the attention effect carries most of the improvement: removing it changes the weighted Brier score by $+15.4\%$ under CPO, $+19.4\%$ under IC and $+21.8\%$ under LASSO, and in the latter two cases \Hyb{w/o\_Attn} is no better than \MBG{}. Under the full covariate set, the contribution is much smaller, at $+4.5\%$, because the linear covariate terms already explain much of the structure that the graph would otherwise supply. The attention effect is therefore most valuable where the covariate representation is deliberately parsimonious, which is also the setting in which a practitioner would most want it. \Hyb{stc}, which adds spatial $k$-nearest-neighbour edges to \Hyb{tc}, behaves differently here from in the simulation. Under the full covariate set, it is the weaker of the two ($-2.2\%$ against $-6.0\%$ in weighted Brier), matching the simulation result that spatial edges are redundant once an SPDE field is present. Under the CPO and IC specifications, however, it is the stronger ($-26.6\%$ against $-20.4\%$, and $-22.9\%$ against $-17.4\%$), and under LASSO the two are equal. Where the covariate set is reduced, the spatial edges appear to recover structure that the smaller covariate set no longer supplies, so the redundancy observed in the simulation does not hold uniformly in this application. The \Hyb{1x} specification provides a further assessment of the hybrid framework under a more parsimonious covariate representation. Despite using only one covariate to define the graph edges, \Hyb{1x} produces substantial improvements relative to \MBG{} across the CPO, IC, and LASSO specifications, suggesting that the hybrid framework can retain its predictive gains under a relatively simple covariate-based graph structure when the attention effect is included. 

The performance of \Hyb{w/o\_SPDE} is not entirely unexpected. Its graph, unlike that of the primary \Hyb{tc} specification, includes spatial edges alongside temporal and covariate-similarity edges (Table~\ref{tab:models}), so spatial information enters the model through graph construction even though the SPDE component is omitted. As a result, part of the spatial dependence that would otherwise be captured through the SPDE component may still be represented by the graph structure, as suggested in Table~\ref{tab:adv_graph} in the simulation study. This provides a plausible explanation for why removing SPDE is less damaging than might otherwise be expected. However, in the simulation study, the outcome has been deliberately generated with a spatial dependence structure that the SPDE is particularly well suited to capture. Therefore, removing the SPDE creates a large loss. But, here, in this malaria application, the observed spatial pattern may be more complex and could be adequately exploited through the spatial edges included in the graph used for the attention effect, even without the explicit SPDE component.

\begin{table}[htbp]
\centering
\caption{Malaria Test-set coverage probability of nominal $95\%$ credible intervals across alternative covariate specifications.}
\label{tab:coverage_covariate}
\footnotesize
\setlength{\tabcolsep}{15pt}
\begin{tabular}{lrrrr}
\toprule
& \textsc{all-cov} & \textsc{cpo} & \textsc{ic}& \textsc{lasso}\\
\midrule
\MBG{} & $0.691$ & $0.635$ & $0.639$ & $0.656$ \\
\Hyb{c} & $0.729$ & $0.750$ & $0.753$ & $0.736$ \\
\Hyb{tc} & $0.729$ & $0.753$ & $0.744$ & $0.732$ \\
\Hyb{stc} & $0.708$ & $0.757$ & $0.750$ & $0.717$ \\
\Hyb{1x} & $0.744$ & $0.748$ & $0.745$ & $0.750$ \\
\Hyb{w/o SPDE} & $0.694$ & $0.736$ & $0.734$ & $0.716$ \\
\Hyb{w/o Attn} & $0.676$ & $0.663$ & $0.634$ & $0.640$ \\
\Hyb{w/o offset} & $0.752$ & $0.749$ & $0.755$ & $0.748$ \\
GAT-KRI & $0.190$ & $0.191$ & $0.191$ & $0.191$ \\
\bottomrule
\end{tabular}
\end{table}

The GATv2 offset is again not so necessary for the predictive gains. Removing it does not weaken the hybrid model; in fact, \Hyb{w/o\_offset} is among the strongest specifications across the different covariate choices. This reinforces the simulation ablation result that the nonlinear structure represented by the offset is largely redundant with the attention field and covariate effects. Residual kriging after GATv2 remains worse than \MBG{} under every covariate specification and also shows weaker uncertainty quantification in this setting (Table~\ref{tab:coverage_covariate}). This differs from the simulation results, in which the same approach produced better uncertainty coverage, possibly due to a different representation of the space--time structure. The empirical variograms here are computed using STIDF, whereas the simulation uses STFDF. This difference may affect how the space--time dependence is estimated and, consequently, the resulting kriging uncertainty.

The attention effect, on the other hand, is generally associated with higher coverage in the hybrid models than in \MBG{} across all four covariate specifications. Coverage increases from approximately $0.635$--$0.691$ for \MBG{} to $0.694$--$0.757$, indicating that the inclusion of the attention effect is associated with improved uncertainty coverage. Across the covariate specifications, \Hyb{tc}, which includes the attention effect, consistently has higher coverage than \Hyb{w/o\_Attn}, where the attention effect is removed. Averaged over the four covariate specifications, \Hyb{tc} covers at $0.740$ against $0.653$ for \Hyb{w/o\_Attn}, an increase of approximately $0.09$.(Table~\ref{tab:coverage_covariate})

\begin{table}[htbp]
\centering
\caption{Posterior mean of $\beta$, quantifying the strength of dependence of the attention effect across different covariate specifications and spatial folds. Values are reported as the mean of the posterior means across spatial folds, with the standard deviation of the posterior means shown in parentheses.}
\label{tab:Beta_for_attn}
\footnotesize
\setlength{\tabcolsep}{8pt}
\begin{tabular}{lrrrr}
\toprule
 & \textsc{all-cov} & \textsc{cpo} & \textsc{ic}& \textsc{lasso}\\
\midrule
\Hyb{c} & 0.739 (0.142) & 0.310 (0.023) & 0.564 (0.061) & 0.707 (0.136) \\
\Hyb{tc} & 0.576 (0.114) & 0.187 (0.135) & 0.395 (0.210) & 0.527 (0.314) \\
\Hyb{stc} & 0.894 (0.129) & 0.270 (0.123) & 0.749 (0.270) & 0.879 (0.094) \\
\Hyb{1x} & 0.065 (0.008) & 0.068 (0.010) & 0.060 (0.004) & 0.059 (0.004) \\
\Hyb{w/o offset} & 0.611 (0.099) & 0.200 (0.139) & 0.486 (0.169) & 0.640 (0.196) \\
\Hyb{w/o SPDE} & 0.962 (0.046) & 0.996 (0.001) & 0.996 (0.003) & 0.983 (0.014) \\
\bottomrule
\end{tabular}
\end{table}

The posterior estimates of the attention dependence parameter in Table~\ref{tab:Beta_for_attn} show how much of the latent dependence is carried by the attention graph rather than by an independent node effect. For the models whose graph uses the full covariate set to define similarity edges, the posterior mass sits well above the prior median of $0.5$ in the \textsc{all-cov} specification: $\beta = 0.739$ for \Hyb{c}, $0.576$ for \Hyb{tc} and $0.894$ for \Hyb{stc}. The malaria data therefore support substantial relational dependence, and the graph is doing more than supplying an unstructured node effect. Two features of the table qualify this. First, $\beta$ is consistently smaller under the CPO specification ($0.187$--$0.310$ for the same three models), where only two covariates define the graph and the similarity structure is correspondingly coarser. Second, \Hyb{1x} holds $\beta$ near $0.06$ throughout, which is consistent with a single-covariate graph carrying little relational information even though that model predicts well; its gains come from the covariate and spatial terms rather than from graph smoothing. The clearest case is \Hyb{w/o SPDE}, where $\beta$ reaches $0.962$--$0.996$: with no spatial field in the model, the attention effect takes on the smoothing role that the SPDE would otherwise perform. Read together with Table~\ref{tab:covariate_comparison}, the dependence parameter and the predictive gain move in the same direction, so the attention component is supported both by the fitted dependence structure and by out-of-sample performance.

\begin{figure}[t]
    \centering
    \includegraphics[width=1\linewidth]{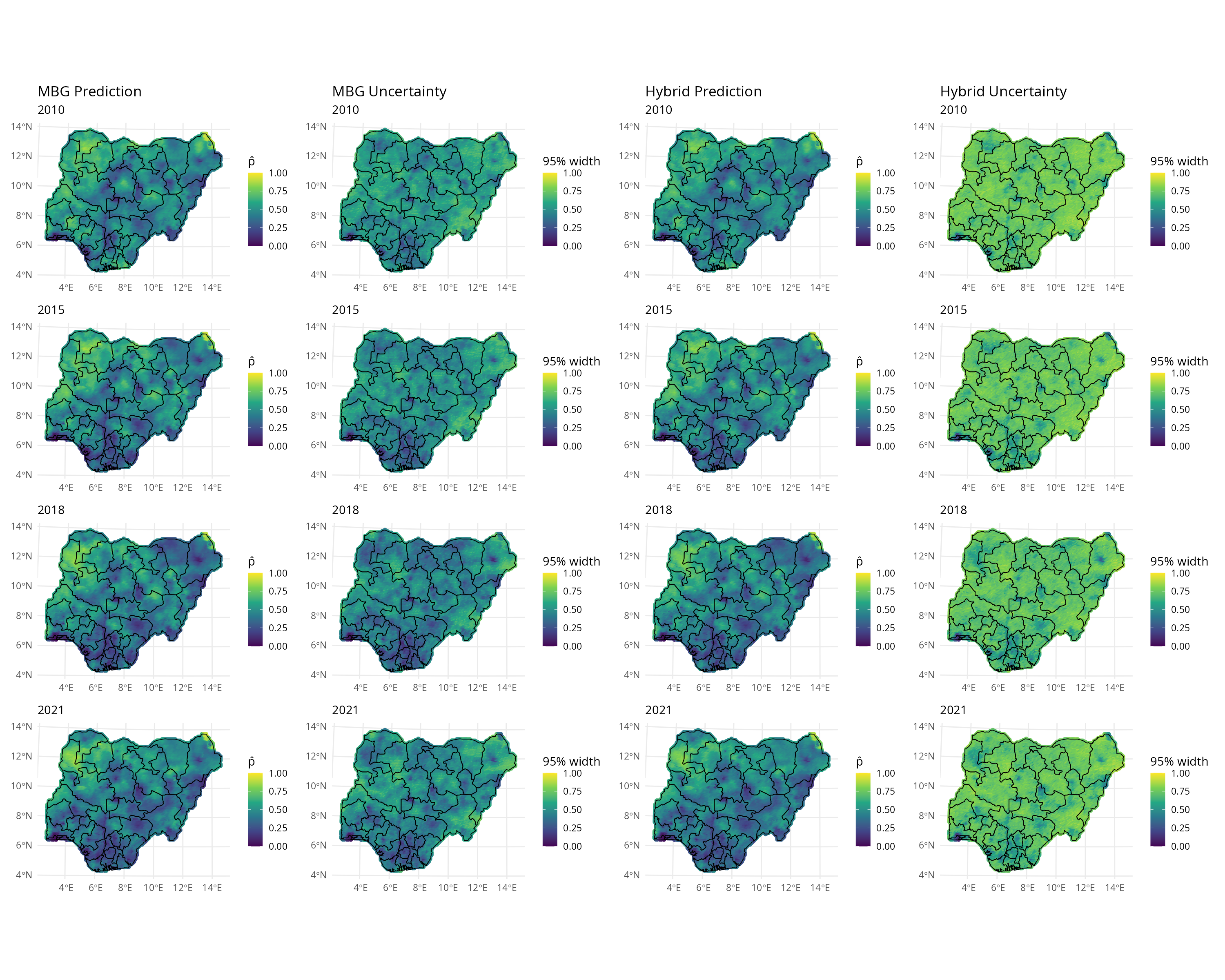}
    \caption{Map of Nigeria showing predicted malaria prevalence from \MBG{} and \Hyb{w/o offset}, with the width of the corresponding $95\%$ posterior credible interval for each}.
    \label{fig:pred_Prevalence}
\end{figure}

Following cross-validation, \MBG{} and \Hyb{w/o offset} were refitted using the full dataset and all the covariates, and their predictions were subsequently mapped onto a 10\,km resolution grid. (Fig.~\ref{fig:pred_Prevalence}), with the latter model selected based on its more consistent predictive performance and coverage among the \Hyb{}s across both the simulation and malaria studies.

\section{Discussion}
The central finding of this study is that the attention random effect introduces dependence that the spatial component of the model does not capture. Using a GATv2-derived Laplacian gives the attention random effect a different role from a conventional spatial random effect. Rather than smoothing only over geographic neighbours, the attention field shares information along edges whose weights are learned from the graph neural network, so that locations are allowed to inform one another when the network judges them relationally similar, even when they are far apart. The construction sits within a well-established family. Equation~\eqref{eq:precision_matrix} is the mixed spatial model of \citet{leroux2000estimation}, in which a mixing parameter interpolates between an independent effect and an intrinsic Gaussian Markov random field \citep{besag1991bayesian,rue2005gaussian,lawson2018bayesian}, and it is close in spirit to the scaled parameterisation of \citet{riebler2016intuitive}. What differs here is the origin of the neighbourhood structure: the weights are estimated from data by an attention mechanism rather than fixed in advance by geography or administrative boundaries. The SPDE field retains its usual role of representing smooth, distance-decaying dependence through a Mat\'ern process \citep{lindgren2011explicit,lindgren2015bayesian}.

This matters because the dependence in many environmental and epidemiological processes is not well described by a function of Euclidean distance alone. Considerable effort has gone into relaxing that assumption within the covariance-function tradition, for example through nonseparable and product-sum space-time families \citep{cressie1999classes,gneiting2002nonseparable,de2001product,cressie2011statistics}, but these constructions still take distance and time lag as their primitives. A learned graph offers a different route: it allows the analyst to specify what similarity should mean, in terms of covariates or other relations, and then to estimate how strongly that similarity governs the latent field. This connects the present work to a growing literature that combines neural architectures with process-based or geostatistical models in the earth and environmental sciences \citep{reichstein2019deep,chen2024deepkriging,hengl2018random,shen2024spatial}, and in particular to graph neural network approaches to small-area estimation \citep{liu2025graph}. The contribution here is to keep the graph inside a fully probabilistic latent Gaussian model, so that the relational structure is one component of a joint posterior rather than a separate preprocessing step.

The simulation results indicate that the two components can be separated, but only when the graph carries information that is not simply a proxy for geographic distance. Identifiability of competing structured effects cannot be assumed, and is known to be delicate in hierarchical spatial models \citep{tang2021identifiability,hodges2010adding,paciorek2010importance,paradinas2023understanding}, so we assessed it directly. Three lines of evidence support it here. First, the posterior scale of the attention field is stable under substantial changes to its hyperprior, moving only from $0.506$ to $0.474$ as the prior expectation of $\operatorname{sd}(A)$ is reduced tenfold, and the posterior scale of the SPDE field is essentially unaffected throughout. If the two effects were competing for the same residual variation, shrinking one would inflate the other. Second, the fitted fields load on different truths: the attention field correlates with the simulated similarity component at $0.74$ to $0.81$ and with the distance component at about $0.15$, while the SPDE field shows the opposite pattern. Third, and most directly, the SPDE field recovers the distance component equally well whether or not the attention effect is present in the model.

The distance-aligned graph experiment shows that the model formulation alone does not guarantee this separation. When the graph is replaced by a purely spatial $k$-nearest-neighbour graph, the correlation between the fitted SPDE field and the true spatial component falls from about $0.90$ to between $0.82$ and $0.84$, and the fitted SPDE standard deviation contracts by roughly $40\%$, indicating that the attention field has begun to absorb variation that belongs to the spatial term. This is a deliberate negative control, and it behaves as the spatial confounding literature would predict \citep{hodges2010adding,paciorek2010importance}. The practical implication is specific rather than general: the attention field earns its place by supplying a different geometry of dependence, and a graph built from geographic proximity supplies no such thing. The same reasoning explains the ablation ordering. Adding spatial edges to the graph degrades performance by about $3\%$, because the SPDE term already represents short-range spatial smoothness, whereas covariate-similarity edges connect locations that are environmentally alike but geographically distant, which is structure the Mat\'ern field cannot express.

The comparison with an unstructured node effect isolates the role of the graph itself. If the improvement came only from adding another random effect, an independent node-level term with the same prior would perform comparably. It does not. Replacing the graph-structured term with an independent one costs between $21$ and $27\%$ in weighted Brier score in every scenario containing a similarity field, in all $100$ replicates, and leaves coverage between $0.52$ and $0.67$, no better than \MBG{}. The reason is structural rather than numerical: an independent effect can absorb variation at observed nodes but transmits nothing to a held-out node, whereas a graph-structured effect allows a held-out node to borrow from its observed neighbours. The gain therefore comes from structured sharing of information, not from added flexibility.

The comparison between the learned attention Laplacian and a fixed covariate $k$-nearest-neighbour graph is more equivocal, and we report it as such. The fixed comparator is slightly ahead under both priors examined, by $0.28\%$ and $0.47\%$ pooled across the nine scenarios, differences that are statistically detectable at $n=900$ but negligible in practical terms and smaller than the $3\%$ penalty incurred by adding spatial edges. We do not read this as evidence that learned weighting is unimportant, because the two graphs differ in two respects at once: the fixed graph uses binary edges and omits the temporal edge block, which the ablation shows is itself worth about $1.2\%$. A clean test would vary edge weighting alone, holding topology fixed, and we did not run one. What the present evidence supports is the weaker and more defensible claim that the graph's composition matters more than whether its weights are learned. This is a useful finding in itself, since a fixed covariate graph is cheaper to construct and easier to interpret.

The malaria application behaves consistently with the simulation. The posterior dependence parameter is substantial under the specifications whose graphs use the full covariate set, reaching $0.739$ for \Hyb{c} and $0.894$ for \Hyb{stc}, so the Nigerian data do support relational dependence of the kind the model is designed to represent. The size of the predictive gain, however, depends on how much of that structure the covariates already carry. Under the full covariate set, removing the attention effect costs only about $4.5\%$ in weighted Brier score, whereas under the parsimonious CPO, IC and LASSO specifications it costs between $15\%$ and $22\%$. Environmental covariates derived from remote sensing are themselves strong predictors of malaria risk \citep{machault2011use,weiss2019mapping}, so this pattern is expected: when the linear predictor is rich, the graph has less left to explain. The practical reading is that the attention effect is most valuable precisely where a parsimonious covariate specification is preferred, whether for interpretability or because covariate data are limited.

Several limitations remain. Coverage improves substantially but still falls short of the nominal $95\%$ level in the application, at roughly $0.71$ to $0.76$ against $0.64$ to $0.69$ for \MBG{}, so the intervals remain too narrow. The malaria analysis does not incorporate DHS survey weights, which may affect population-level inference, and the deliberate displacement of DHS cluster coordinates affects both covariate extraction and the representation of short-range spatial dependence. Uncertainty in the learned graph is also not propagated into the Bayesian stage: Monte Carlo dropout is used to average the attention coefficients, but the resulting Laplacian is then treated as fixed, so the reported posteriors condition on an estimated graph. Treating the graph as a random object, for example by placing a prior over edge weights or by propagating the dropout ensemble through the second stage, is the natural next step and would likely widen intervals in the direction the coverage results suggest they should move.

\section{Conclusion}
This study developed a hybrid spatial modelling framework that combines distance--based spatial dependence with attention--based relational dependence. The results show that these two components provide complementary information: the spatial field accounts for dependence associated with geographic distance, while the graph-structured component represents association along a covariate-similarity neighbourhood. The contribution of this component is not simply due to adding another random effect, since an unstructured effect of the same prior recovers none of the gain; its value comes from the structure of the neighbourhood over which information is shared. Our experiments do not, however, support attributing the gain specifically to the attention weights themselves, since a fixed covariate graph performs comparably. In addition to improving predictive performance, the attention component improves the representation of predictive uncertainty coverage. The framework therefore provides a flexible way to incorporate multiple forms of dependence within a statistically structured spatial model, while retaining the ability to use more parsimonious covariate representations. These results validate the use of hybrid models as a statistically grounded alternative for spatial epidemiology and public health decision--making, and recommend further refinement of uncertainty calibration.

\section*{Data and code availability}
All models were fitted with \texttt{R--INLA} via \texttt{inlabru}; the GATv2 component was fitted in \texttt{PyTorch Geometric}, with the two stages linked through file exchange. The malaria prevalence data are publicly available from the Demographic and Health Surveys programme \citep{NigeriaDHS_Grouped}. All code used for the simulation study and the malaria application is maintained in the GitHub repository \url{https://github.com/btoba37/Hybrid_Graph_Attention_Geostatistical_Methods}; as this is currently a live repository rather than a permanently archived snapshot, a Zenodo DOI will be minted for the accepted version so that this statement points to an immutable, citable archive.

\clearpage
\appendix
\section{Supplementary Figures and Tables}\label{app:supplementary}
\begin{longtable}{>{\raggedright\arraybackslash}p{3.5cm} >{\raggedright\arraybackslash}p{9.3cm}}
\caption{Definition of Covariates}
\label{tab:def. of covs} \\
\hline
\textbf{Covariates} & \textbf{Definition} \\
\hline
\endfirsthead

\hline
\textbf{Covariates} & \textbf{Definition} \\
\hline
\endhead
\hline
\endfoot

\hline
\endlastfoot

Temperature & Air temperature measured at height of 2 meters. \\
Min Temperature & Lowest air temperature at height of 2 meters. \\
Max Temperature & Highest air temperature at height of 2 meters. \\
Skin Temperature & Temperature of the Earth’s surface. \\
Soil Temperature & Temperature of the upper soil layer spanning 0–7 cm depth, measured at the midpoint. \\
Volumetric Soil Water & The volume of water present in the top soil layer (0–7 cm). \\
Relative Humidity & Measures how close the air is to total saturation with water vapor at the current temperature. \\
Surface Runoff & Accumulated depth of water flowing over the land surface when soil infiltration capacity is exceeded. \\
Sub--surface Runoff & Accumulated depth of water draining laterally through the soil beneath the surface. \\
Total Evaporation & Combined evaporation and plant transpiration to the atmosphere. \\
Total Precipitation & Accumulated rain and snow, excluding fog, dew, and evaporated precipitation. \\
Potential Evaporation & A theoretical calculation of evaporation rates assuming the surface is agricultural land with an unlimited water supply. \\
Surface Net Solar Radiation & The actual amount of shortwave solar energy absorbed by the ground. \\
EVI & A metric that assesses vegetation greenness. \\
NDVI & A widely used ratio that indicates the density and health of live vegetation. \\
Distance to Water & The calculated distance (in meters) from a land pixel to the nearest pixel identified as a water body. \\
Elevation & Terrain's vertical height relative to mean sea level. \\
Slope & The degree of inclination of the terrain surface. \\

\end{longtable}

\begin{flushleft}
\footnotesize \textit{Note}: All covariates are extracted at a spatial resolution of 10--km from Google Earth Engine.
\end{flushleft}

\begin{table}[htbp]
\centering
\caption{Percentage change in test-set Binomial Log Score and Weighted Brier score relative to \MBG{}, simulation block hold-out, $100$ replicates. Negative values are better
than \MBG{}.}
\label{tab:relative_scores_block}
\footnotesize
\setlength{\tabcolsep}{4.5pt}
\begin{tabular}{lcccccccc}
\toprule
& \Hyb{c} & \Hyb{tc} & \Hyb{stc} & \Hyb{1x} &
\Hyb{w/o SPDE} & \Hyb{w/o Attn} & \Hyb{w/o off} & GAT-KRI \\
\midrule

\multicolumn{9}{l}{\textit{Binomial Log Score}} \\
s1 & $+0.2$ & $+0.6$ & $+3.8$ & $+0.1$ & $+35.3$ & $+0.4$ & $-0.0$ & $+43.6$ \\
s2 & $-11.6$ & $-10.9$ & $-7.8$ & $-11.2$ & $+3.6$ & $+0.8$ & $-11.7$ & $+13.3$ \\
s3 & $-11.5$ & $-10.2$ & $-5.2$ & $-10.2$ & $+4.2$ & $+1.1$ & $-11.7$ & $+11.0$ \\
s4 & $-11.1$ & $-10.1$ & $-4.9$ & $-10.2$ & $+5.4$ & $+0.7$ & $-11.3$ & $+15.1$ \\
s5 & $-11.1$ & $-10.3$ & $-5.3$ & $-10.8$ & $+7.6$ & $+0.7$ & $-11.5$ & $+13.4$ \\
s6 & $-16.1$ & $-15.5$ & $-15.3$ & $-14.4$ & $-15.2$ & $-0.1$ & $-16.0$ & $+5.1$ \\
s7 & $-2.2$ & $-0.8$ & $+3.4$ & $-1.1$ & $+28.3$ & $+0.8$ & $-1.6$ & $+40.0$ \\
s8 & $-10.3$ & $-9.6$ & $-5.6$ & $-9.9$ & $+7.6$ & $+0.7$ & $-10.4$ & $+16.4$ \\
s9 & $-10.0$ & $-9.2$ & $-6.5$ & $-9.7$ & $+4.7$ & $+0.8$ & $-10.4$ & $+17.2$ \\

\midrule
\multicolumn{9}{l}{\textit{Weighted Brier}} \\
s1 & $+0.6$ & $+1.9$ & $+14.3$ & $+0.1$ & $+147.9$ & $+1.6$ & $-0.2$ & $+180.7$ \\
s2 & $-18.7$ & $-17.6$ & $-12.4$ & $-17.9$ & $+7.4$ & $+1.2$ & $-19.1$ & $+24.6$ \\
s3 & $-18.6$ & $-16.3$ & $-7.6$ & $-16.3$ & $+8.6$ & $+1.7$ & $-18.9$ & $+20.4$ \\
s4 & $-18.4$ & $-16.6$ & $-7.6$ & $-16.9$ & $+10.5$ & $+1.2$ & $-18.8$ & $+27.3$ \\
s5 & $-17.8$ & $-16.4$ & $-8.0$ & $-17.1$ & $+14.5$ & $+1.2$ & $-18.4$ & $+25.4$ \\
s6 & $-29.3$ & $-28.4$ & $-28.0$ & $-26.3$ & $-27.9$ & $-0.3$ & $-29.2$ & $+11.2$ \\
s7 & $-6.0$ & $-1.7$ & $+12.6$ & $-2.7$ & $+99.0$ & $+2.5$ & $-4.2$ & $+136.0$ \\
s8 & $-17.3$ & $-16.1$ & $-8.6$ & $-16.6$ & $+15.6$ & $+1.3$ & $-17.5$ & $+31.9$ \\
s9 & $-16.5$ & $-15.2$ & $-10.3$ & $-16.1$ & $+10.2$ & $+1.5$ & $-17.2$ & $+33.5$ \\

\bottomrule
\end{tabular}
\end{table}

\begin{table}[htbp]
\centering
\caption{Empirical coverage of nominal $95\%$ credible intervals for the
latent risk $p_i$, test set, simulation block hold-out, $100$ replicates.}
\label{tab:block_coverage}
\footnotesize
\begin{tabular}{lrrrrrrrrr}
\toprule
& s1 & s2 & s3 & s4 & s5 & s6 & s7 & s8 & s9 \\
\midrule
\MBG{}           & $0.938$ & $0.748$ & $0.734$ & $0.749$ & $0.731$ & $0.490$ & $0.913$ & $0.753$ & $0.749$ \\
\Hyb{c}          & $0.951$ & $0.938$ & $0.930$ & $0.938$ & $0.928$ & $0.938$ & $0.946$ & $0.935$ & $0.934$ \\
\Hyb{tc}         & $0.946$ & $0.935$ & $0.926$ & $0.932$ & $0.924$ & $0.938$ & $0.940$ & $0.929$ & $0.929$ \\
\Hyb{stc}        & $0.913$ & $0.918$ & $0.897$ & $0.907$ & $0.901$ & $0.936$ & $0.905$ & $0.909$ & $0.911$ \\
\Hyb{1x}         & $0.951$ & $0.936$ & $0.928$ & $0.934$ & $0.930$ & $0.934$ & $0.944$ & $0.931$ & $0.932$ \\
\Hyb{w/o SPDE}   & $0.793$ & $0.891$ & $0.873$ & $0.882$ & $0.871$ & $0.936$ & $0.818$ & $0.879$ & $0.885$ \\
\Hyb{w/o Attn}   & $0.935$ & $0.745$ & $0.723$ & $0.742$ & $0.722$ & $0.512$ & $0.904$ & $0.746$ & $0.745$ \\
\Hyb{w/o offset} & $0.952$ & $0.940$ & $0.932$ & $0.939$ & $0.932$ & $0.940$ & $0.947$ & $0.935$ & $0.935$ \\
GAT-KRI & $0.935$ & $0.932$ & $0.927$ & $0.927$ & $0.924$ & $0.944$ & $0.937$ & $0.928$ & $0.929$ \\
\bottomrule
\end{tabular}
\end{table}

\begin{figure}[t]
    \centering
    \includegraphics[width=0.9\linewidth]{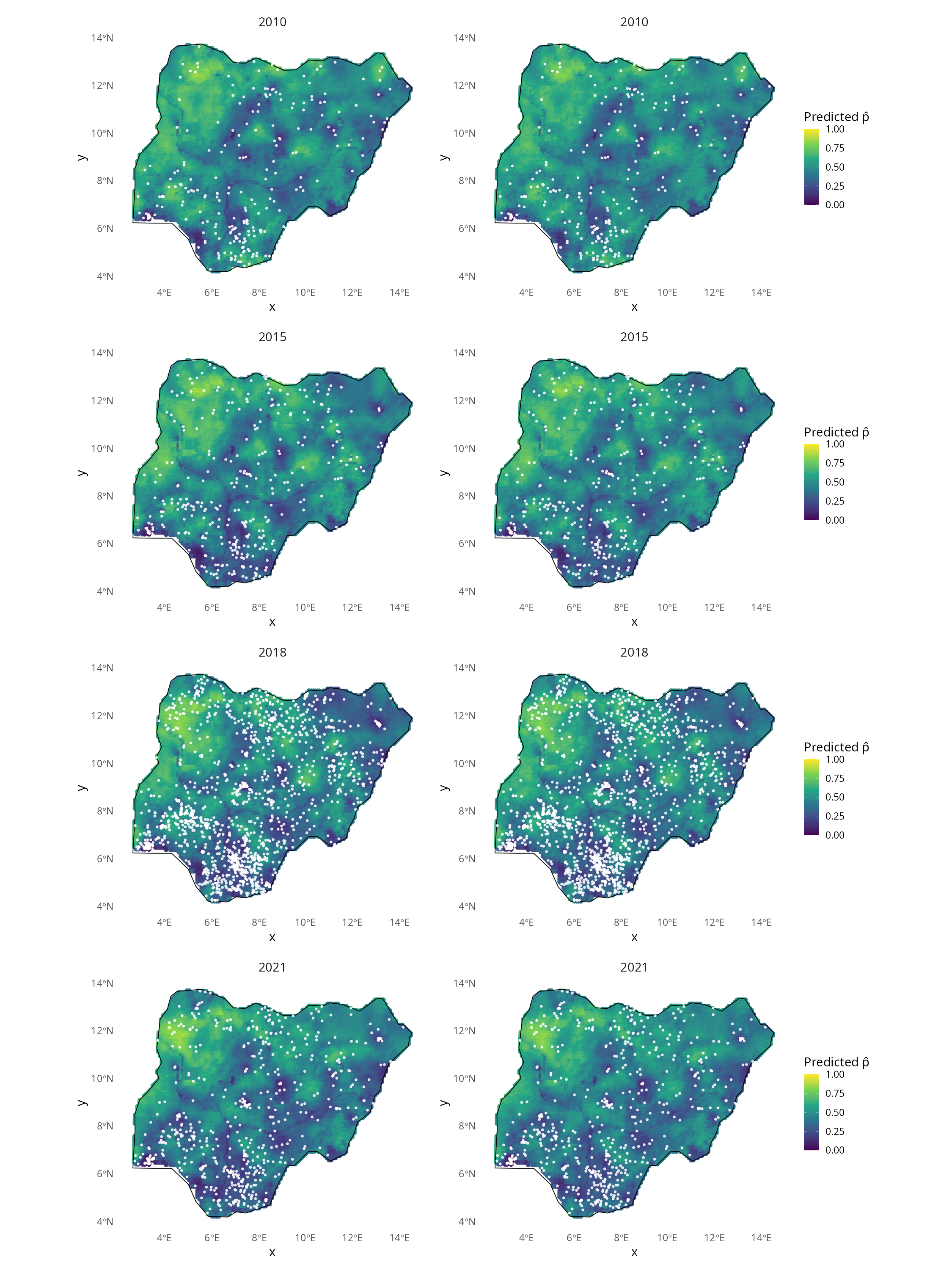}
    \caption{Map of Nigeria showing predictive surface with observed data points. First column predicted by \MBG{} and the second by \Hyb{}}
    \label{fig:pred_Prevalence_plus_obs}
\end{figure}

\begin{figure}[htbp]
\centering
\includegraphics[width=0.9\textwidth]{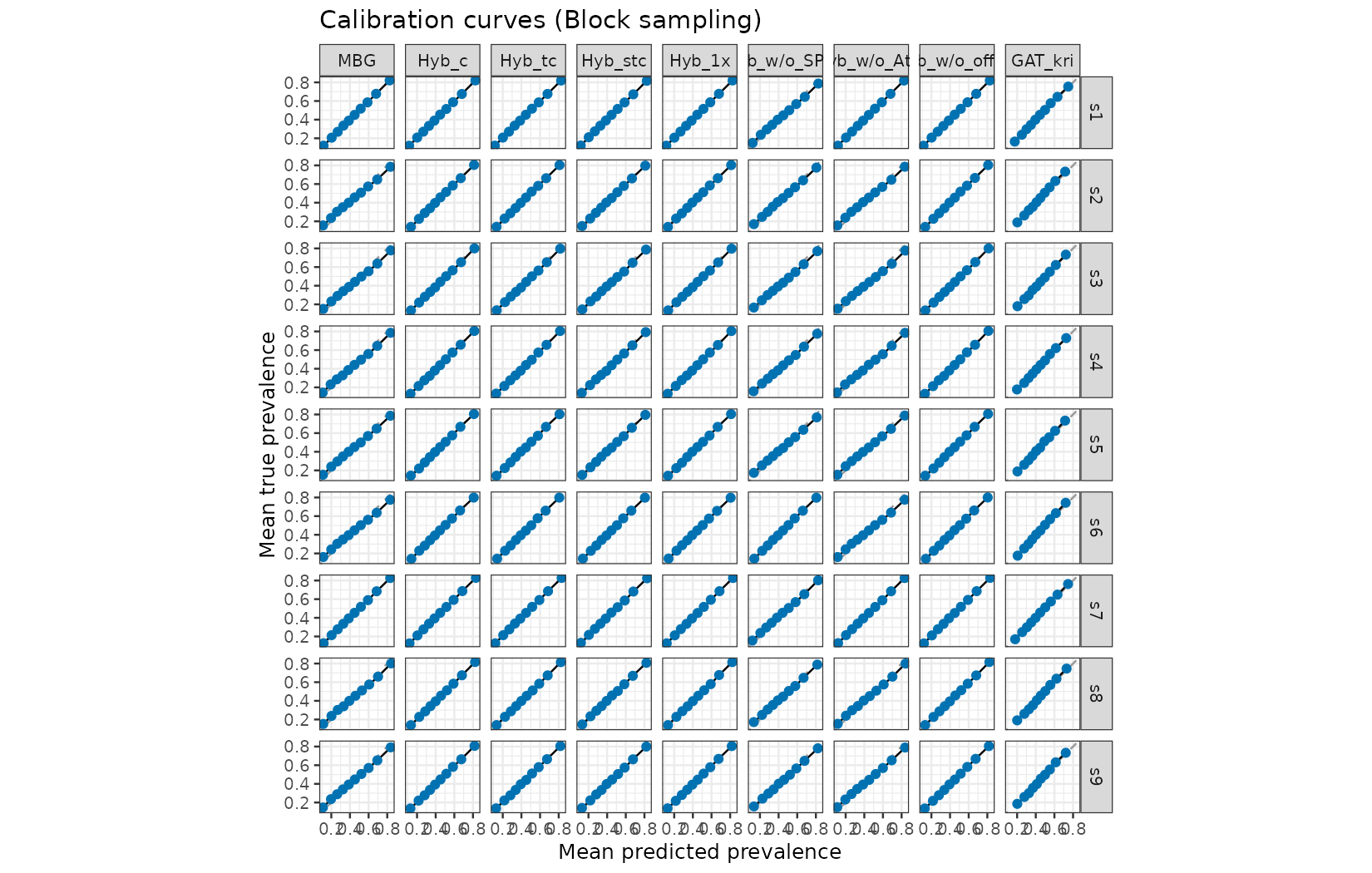}
\caption{Calibration by decile of predicted risk, test set, block hold-out, pooled over $100$ replicates. Rows are scenarios, columns are models; the dashed line is perfect calibration.}
\label{fig:block_calib}
\end{figure}

\clearpage
\bibliographystyle{elsarticle-harv} 
\bibliography{references}
\end{document}